\documentclass[aps,prl,reprint,superscriptaddress,floatfix]{revtex4-2}

\usepackage[left]{lineno} 
\usepackage{color,soul}
\usepackage{amssymb,amsmath,amsfonts}
\usepackage{graphicx}
\usepackage{dcolumn}
\usepackage{multirow}
\usepackage{hyperref}
\usepackage[dvipsnames]{xcolor}
\usepackage{siunitx}
\usepackage[T1]{fontenc}
\usepackage{natbib}

\begin{document}
\title{Cross-hatch strain effects on SiGe quantum dots for qubit variability estimation} 

\author{Luis Fabián Peña}
\affiliation{Sandia National Laboratories, Albuquerque NM, USA}
\altaffiliation{Department of Physics, Baylor University, Waco, TX 76798, USA}

\author{Mitchell I. Brickson}
\affiliation{\mbox{Center for Computing Research, Sandia National Laboratories, Albuquerque NM, USA}}

\author{Fabrizio Rovaris}
\affiliation{Department of Materials Science, University of Milano-Bicocca, Milano, Italy}

\author{J. Houston Dycus}
\affiliation{Advanced Microscopy, Eurofins EAG Materials Science, Raleigh NC, USA}

\author{Anthony McDonald}
\affiliation{Sandia National Laboratories, Albuquerque NM, USA}

\author{Zachary T. Piontkowski}
\affiliation{Sandia National Laboratories, Albuquerque NM, USA}

\author{Joel Benjamin Ruzindana}
\affiliation{\mbox{Department of Chemistry and Physics, University of Arkansas at Pine Bluff, Pine Bluff AR, USA}}

\author{Adelaide M. Bradicich}
\affiliation{\mbox{Center for Integrated Nanotechnologies, Sandia National Laboratories, Albuquerque NM, USA}}

\author{Don Bethke}
\affiliation{\mbox{Center for Integrated Nanotechnologies, Sandia National Laboratories, Albuquerque NM, USA}}

\author{Robin Scott}
\affiliation{\mbox{Lawrence Semiconductor Research Laboratory, Inc., Tempe AZ, USA}}

\author{Thomas E. Beechem}
\affiliation{\mbox{Mechanical Engineering and Birck Nanotechnology Center, Purdue University, West Lafayette IN, USA}}

\author{Francesco Montalenti}
\affiliation{Department of Materials Science, University of Milano-Bicocca, Milano, Italy}

\author{N. Tobias Jacobson}
\email{ntjacob@sandia.gov}
\affiliation{\mbox{Center for Computing Research, Sandia National Laboratories, Albuquerque NM, USA}}

\author{Ezra Bussmann}
\email{ebussma@sandia.gov}
\affiliation{Sandia National Laboratories, Albuquerque NM, USA}

\date{\today}



\begin{abstract}
SiGe heterostructures integrated with Si via virtual substrate (VS) growth are promising hosts for spin qubits. While VS growth targets plastic relaxation, residual cross-hatch strain inhomogeneity propagates into heterostructure overgrowth. To quantify strain inhomogeneity's influence on interface structure and qubit properties, we measure strained-silicon (s-Si)/Si$_{0.7}$Ge$_{0.3}$ heterostructures on 25 wafers processed via standard commercial chemical vapor deposition. Spatially-aligned images of strain (Raman microscopy) and interface structure (atomic force microscopy and cross-sectional scanning transmission electron microscopy) reveal strain–roughness interplay. A strain-driven surface diffusion model predicts the roughness and its temperature dependence. Measured strains suggest spurious double-dot qubit detunings of 0.1 meV over 100 nm distances may result. Modeling shows that interface roughness (atomic steps), when convolved with alloy disorder, only modestly reduces valley splitting (70$\pm$13 \textit{vs.} 77$\pm$14 $\mu$eV on average). Our findings point to thicker VS buffer layers beneath heterostructures and lower-temperature growth (T $\le$ 700 $^{\circ}$C) to limit roughening.

\end{abstract}

\maketitle{}
  
\section{Introduction}

SiGe heterostructures on virtual substrates (VSs) are a leading approach for integrating gate-defined electron or hole quantum dot spin qubits onto Si~\cite{Currie1998, Sawano2003, Zwanenburg2013, Burkhard2023, richardson2016, Deelman2016, Scappucci2021, Kawakami2014, zajac2016,  Xue2022,mills2022,noiri2022,Weinstein2023}. For electron confinement, tensile strained-Si (s-Si) layers act as 100-200 meV-deep quantum wells between relaxed Si$_{1-x}$Ge$_{x\sim0.3\pm0.1}$ layers, Fig. \ref{Fig1}~a~\cite{Zwanenburg2013,Burkhard2023, richardson2016, Deelman2016, Scappucci2021}.
Meanwhile, the VS effects a metamorphic transition between relaxed single-crystal Si$_{0.7}$Ge$_{0.3}$ and the Si wafer, yielding materials quality required to fabricate qubits~\cite{richardson2016, Deelman2016, Neyens2024, Koch2025}. The VS approach is key to leveraging existing Si manufacturing for fabricating qubits in SiGe heterostructures using chemical vapor deposition (CVD), the industry's preferred method, with few alternatives, e.g., single-crystal SiGe wafers or  membranes~\cite{Weinstein2023, Neyens2024, Koch2025, Yonenaga2005, Subramanian2025, Paskiewicz2011}. Paralleling conventional Si metal oxide semiconductor (SiMOS \textit{sans} Ge) qubit development, SiGe processing with advanced-node manufacturing has reached high-yield qubits with competitive coherence times and high-fidelity operations at \textit{Intel Corp.} and \textit{imec} (Belgium) \cite{Kozlowski2013, Neyens2024,Koch2025, Steinacker2024,Steinacker2025}. 

SiGe heterostructures confine qubit spins in entirely crystalline environments, unlike SiMOS interfaces, with favorably lower disorder than SiMOS as inferred from proxy metrics including lower metal-insulator percolation threshold density and higher electron mobility~\cite{Zwanenburg2013, Burkhard2023, Lu2011, Lawrie2020, gyure2021, Scappucci2021}. But, other materials disorder unique to SiGe epitaxy points to appreciable qubit variability, which may be a hurdle to scaling up SiGe technology~\cite{Boykin2004,Boykin2004c, Friesen2010, gyure2021, Scappucci2021,Martinez2022, Neyens2024, Marcks2025}. Interface disorder affects spin qubits via mechanisms that couple spin, valley, and orbital degrees of freedom and depend on elemental composition in the vicinity of the quantum dot. This results in, for example, perturbations to exchange interactions between electron spins that drive single- and two-qubit logic gates. Some variability results from Si-Ge alloy disorder and intermixing \cite{jernigan1996,Uberuaga2000,Qin2000,Boguslawski2002,Hannon2004,Akis2005,Zipoli2008,Neyens2018,Esposti2024}, non-conformal epitaxy causing Si QW thickness variations \cite{Pena2024, PaqueletWuetz2022}, and inhomogeneous strain due to VS plastic relaxation~\cite{Fitzgerald1997, Hsu1992,10.1063/1.1629142}. These effects must also be considered alongside extrinsic effects, e.g., from the gate stack, including the gate dielectric \cite{Martinez2024} and metal electrode-induced strains~\cite{Park2016, Corley-Wiciak2023}.

Virtual substrate materials mediate a plastic relaxation of 1.2\% between the Si wafer and Si$_{0.7}$Ge$_{0.3}$, aiming to maximize relaxation and minimize threading dislocation density~\cite{Deelman2016, richardson2016, Scappucci2021}. Typically, plastic relaxation is driven by depositing layers with either smooth or step-wise increases in Ge$_{x}$ content, normally around $\Delta x\sim0.1/\mu$m, at high temperature ($>$750 \textdegree C). This approach enhances misfit dislocation mobility, relaxation per dislocation, and mitigates threading dislocation densities ($<10^{6}$ cm$^{-2}$) through a modified Frank-Read mechanism, as identified by LeGoues \textit{et al.}~\cite{Fitzgerald1991, Legoues1991, Mooney1993, Mooney1995, Currie1998, Sawano2003, richardson2016, Scappucci2021, Pena2024, Esposti2024}. The graded layers are overgrown with relaxed, misfit-free, constant-composition Si$_{0.7}$Ge$_{0.3}$ \cite{Currie1998, Sawano2003}. While VSs reach a state of bulk \textit{mean} relaxation, prior studies indicate the presence of strain inhomogeneity with a salient cross-hatch pattern resulting from misfit dislocation pile-ups distributed nonuniformly in the graded-layer~\cite{Fitzgerald1997, jesson1997, Chen2002, Sawano2003, 10.1063/1.1629142,Kutsukake2004,Sawano2005, Kovalskiy2019}.  
Cross-hatch strains cause appreciable varying local crystal tilts ($0.2$\textdegree) and noticeable growth surface roughening on the scale of 10 nm root mean square (RMS)~\cite{Hsu1992, Lutz1995, Mooney1999, Evans2012, Tilka2016, Zoellner2015, Corley-Wiciak2023,Esposti2024}. Cross-hatching is observed in other metamorphic materials, such as GaAs/InGa(Al)As~\cite{Kovalskiy2019}. For Si spin qubits, cross-hatch surface roughness is incompatible with manufacturing, which requires planar substrates for 3D gate integration~\cite{Neyens2024}. Typically, VSs are finished with planarization, e.g., chemical mechanical polish (CMP), yielding flat surfaces for heterostructure growth (Fig. \ref{Fig1} b)~\cite{Currie1998, Sawano2003}.

While virtual substrate planarization mitigates roughness, it has no apparent effect on cross-hatch strain, leaving intact a strain template that is coherently inherited by subsequent heterostructure layers. This impacts growth and interface structure~\cite{Sawano2003, Zoellner2015, Shida2017}.
Cross-hatch strain and structure in SiGe VSs are commonly measured with nano-scale X-ray diffraction (XRD), reflection (XRR), and Raman microscopy. Studies consistently link cross-hatch strain to the underlying misfit dislocation network and find it largely unaffected by planarization~\cite{Hsu1992, Chen2002,  Sawano2003,  Zoellner2015, Shida2017,  Richardson2017}.
Through Raman microscopy, a few works examined the strain inhomogeneity in layers undergoing plastic relaxation and reported cross-hatch strain fluctuations on the order of 10$^{-3}$
~\cite{Chen2002, Sawano2003, Kutsukake2004,Sawano2005}. 
While  nanoprobe XRD, tomography, and XRR  yield high-accuracy \textit{mean} local structure measurements in the direction normal to the surface, these measurements are usually supplemented with 2D/3D real-space structure images to establish more complete strain-structure relationships.
For VS materials, nanoprobe XRD and Raman strain and tilt data have been correlated with cross-sectional electron microscopy and surface structure [atomic force microscopy (AFM)] via coincident-site measurements~\cite{Chen2002, Sawano2003, Zoellner2015}.

Recent studies link inhomogeneous strain to adverse effects on qubit performance~\cite{Evans2012,Thorbeck2015, Tilka2016, Corley-Wiciak2023}. For example, Evans {\it et al.} used coherent X-ray imaging to map strain in the vicinity of a buried quantum well. Their results indicated a strain difference of about 10$^{-6}$ between the top and bottom of the s-Si quantum well, leading to 14 $\mu$eV variability in the conduction band edge, which is similar to important qubit energy scales such as valley splittings and typical Zeeman energies.  
To the best of our knowledge, such studies have not yet extended to strain-structure inheritance through the qubit heterostructure growth process, or connected cross-hatch strain-structure relationships to putative qubit impacts. The cross-hatch strain fluctuations, (0.001) are comparable to strains induced by metal gate electrodes fabricated on top of the qubit structures~\cite{Thorbeck2015, Park2016, Corley-Wiciak2023}. Strains shift the conduction band valley energies, which has non-negligible effects on the qubit potential energy landscape~\cite{Thorbeck2015, Corley-Wiciak2023}. Sawano {\it et al.} showed that cross-hatch drives surface roughening during growth, through the `memory effect'~\cite{Sawano2003}. Furthermore, Rovaris {\it et al.} directly showed that surface re-roughening results from residual strain fluctuations at the free surface that are present due to the plastic relaxation of the  VS~\cite{RovarisPRB2019}. The resulting nanoscale roughness, which reflects the cross-hatch pattern, introduces new atomic steps that are predicted to appreciably impact the valley splitting in quantum dots~\cite{Friesen2010, Pena2024}. Elucidating the role of cross-hatch strain inhomogeneity in setting single-quantum dot spectral characteristics, e.g., orbital and valley splitting, and inter-dot couplings, e.g., detuning and exchange, is both timely and essential to understand variability in future larger-scale qubit systems or when moving qubits through large disorder landscapes, e.g., during shuttling~\cite{Seidler2022, Volmer2024, Neyens2024, Marcks2025}.

For this purpose, we report growth studies tracking the surface morphology and strain evolution of 25 wafers in a standard commercial CVD process, as illustrated in Fig \ref{Fig1}. Substrate strain inhomogeneity exerts driving forces on surface dynamics during annealing and growth processes. We show that surface roughness is correlated to both the underlying strain fields and buried heterostructure interface morphology. A strain-driven surface diffusion model explains the measured strain-structure relationship, as well as the mean-square roughness, correlation length, and rate of roughening. Importantly, we find that the upper and lower quantum well interfaces are strongly correlated. Moreover, strain inhomogeneity may modify the energetics of double- or triple- quantum dots. We estimate the magnitude of this effect based on deformation potential theory, finding that measured residual strains may induce spurious double-dot detunings of 0.1 meV over 100 nm length scales.
Finally, using measured Si/SiGe interface alloy intermixing lengths and atomic-step structures, we compute quantum dot valley splitting variability through atomistic multivalley effective mass theory. Quantum computation with Si/SiGe quantum dot electron spins requires consistently large valley splitting to suppress errors caused by excitation into other valley-orbital states~\cite{Burkard2021}. For our measured interface alloy profiles, we predict mean valley splittings of approximately 70 $\mu$eV. We find that Si/SiGe interface alloy disorder alone leads to significant valley splitting variability, and that including interface steps causes further modest ($\sim$10\%) reduction in average valley splitting. This relative insensitivity of valley splitting to interface steps in the presence of significant interface intermixing is consistent with previous work~\cite{Losert2023,Pena2024}. We anticipate that such valley splitting statistics may inform future projections of spin qubit device yield.

\section{Results}

\subsection{Experiment: CVD growth surface roughness study}
 The 25 wafers in this study were grown by CVD at Lawrence Semiconductor Research Laboratory, Inc. Detailed characterization focused on the wafers processed for virtual substrate preparation and heterostructure growth at 600 \textdegree C, specifically, wafers \#1–15, as shown in Fig. \ref{Fig1}. The full list of wafer processing steps is provided in Supplementary Tables S1, S2, and Fig. S1. Preliminary results from heterostructure growth at 700 \textdegree C, Supplementary Fig. S1, were consistent with those at 600 \textdegree C, so further investigation at 700 \textdegree C was not pursued. The experiment tracked surface topography and near-surface strain inhomogeneity through growth of a VS and s-Si/SiGe quantum well heterostructure, Fig. \ref{Fig1}~a. Our growth recipe is comparable to others for spin qubit material in prior studies, e.g., [Ref. \cite{Mi2015}].  In our experiment, a wafer was extracted and analyzed after each sequential growth step or new layer, as depicted in Fig. \ref{Fig1}~b-c. Surface structure was measured by AFM. The strain inhomogeneity was mapped using a well-established Raman microscopy imaging technique to track the Si-Si LO$_z$ phonon peak position within the SiGe layers which shifts in response to strain variations, $\delta\epsilon_{||}$ (see Methods and Supplementary Figs. S3 and S4). The interface atomic structure of the complete heterostructure was measured using cross-sectional high-angle annular dark field scanning transmission electron microscopy (HAADF-STEM). For two monitor wafers, the SiGe composition (Si$_{1-x}$Ge$_{x}$, with $x =0.30\pm0.01$) and strain relaxation were measured using secondary-ion mass spectrometry (SIMS) and X-ray diffraction (XRD), respectively, as shown in Supplementary Fig. S2.  

\begin{figure*}
 \includegraphics[width=0.9\textwidth]{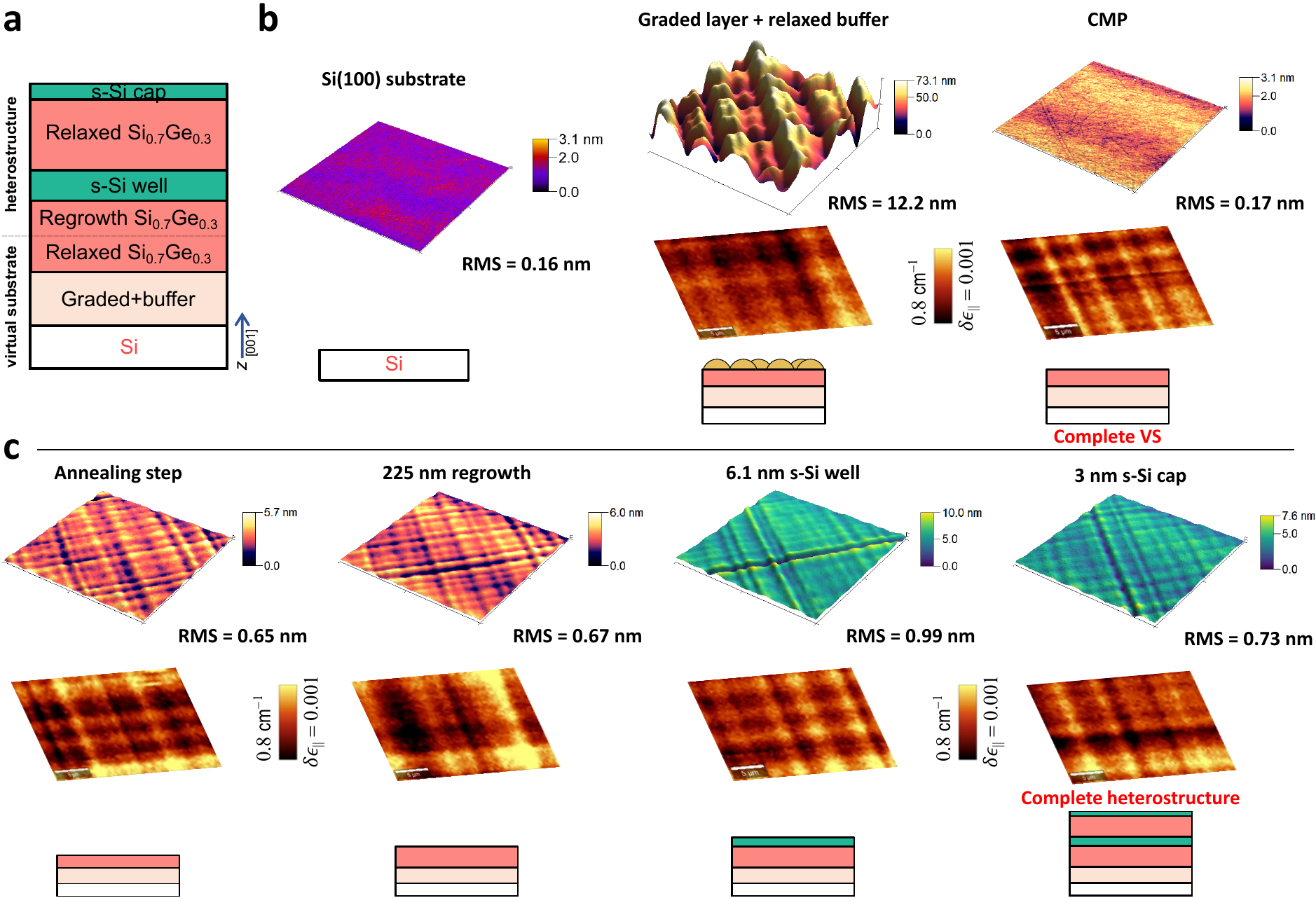}
\caption{\textbf{The growth experiment process showing virtual substrate (VS) preparation and heterostructure growth.} \textbf{a} A schematic cross-section showing the VS and heterostructure layers. \textbf{b} The process flow for the VS with AFM images tracking surface evolution: (wafer \#1) Si(100) substrate, (wafer \#3) relaxed Si$_{0.7}$Ge$_{0.3}$ on graded growth, (wafer \#5) surface post chemical mechanical polishing. \textbf{c} The heterostructure growth with AFM images: (wafer \#7) epitaxy-ready Si$_{0.7}$Ge$_{0.3}$ relaxed buffer (post annealing), (wafer \#13) 225 nm SiGe regrowth, (wafer \#14) 6.1 nm Si QW, and (wafer \#15) 3 nm s-Si cap. Imaging of the relaxed Si$_{0.7}$Ge$_{0.3}$ heterostructure layer was not undertaken; given the thin s-Si cap overlayer, it is assumed to be nominally similar. Heterostructure deposition, T = 600 \textdegree C. Epitaxy-ready surfaces were prepared by dilute HF and H$_2$ anneal at 900 \textdegree C for 2 min prior to heterostructure growth. AFM images show 20 $\mu$m $\times$ 20 $\mu$m area. RMS roughness values were obtained from each processed wafer and from different areas, rather than from a single identical location. Raman microscopy imaging over a 20 $\mu$m $\times$ 15 $\mu$m area, shown above the schematic drawings, reveals the underlying strain inhomogeneity in the layered structures.}
\label{Fig1} 
\end{figure*}

Virtual substrate growth started with deposition of a 3 µm-thick linear graded layer (Ge fraction increased at 0.1/µm) on flat Si(100) with roughness RMS~$=0.16\pm0.02$~nm for wafer \#1, as shown in Fig. \ref{Fig1}~b. The graded layer relaxes completely, on average, during deposition (see XRD data in Supplementary Fig. S2), and is subsequently capped with a relaxed Si$_{0.7}$Ge$_{0.3}$ buffer layer (SIMS data in Supplementary Fig. S2). The graded and relaxed buffer, Fig. \ref{Fig1}~b wafer \#3, had cross-hatch roughness (RMS = 12.2~nm) owing to plastic relaxation of the graded layer. The significant surface roughness complicates qubit fabrication, which demands planar substrates. Therefore, the wafers were planarized using CMP. The CMP yields a flat Si$_{0.7}$Ge$_{0.3}$ virtual substrate for subsequent heterostructure growth with RMS roughness of 0.17~nm, as shown in Fig. \ref{Fig1}~b wafer \#5. Note that while CMP planarizes the surface, it does not eliminate strain inhomogeneity, i.e., fluctuations around the mean lattice constant, that arise from relaxation during graded layer growth\cite{10.1063/1.1629142,Evans2012} (see Fig. \ref{Fig2} b). 

To prepare for heterostructure deposition, each VS-SiGe wafer was cleaned, introduced into the CVD tool, and annealed  (T = 900~\textdegree C/2 min, in H ambient).  Following the anneal, the cross-hatch roughness partially re-emerged (RMS = 0.63 nm), as shown in Fig. \ref{Fig1}~c, wafer \#7. 
Next, heterostructures were deposited at T = 600 \textdegree C. Despite differing layer compositions and strain states (>1\% strained-Si vs. relaxed SiGe), as shown in Fig. \ref{Fig1}~c, all subsequent heterostructure layers for wafers \#7, 13, 14, 15,  have comparable cross-hatch features and RMS roughness from 0.63-0.80 nm (see Supplementary Fig. S1).

\subsection{Experiment: Cross-correlating interface structure.}

\begin{figure*}
\centering
\includegraphics[width=0.9\textwidth]{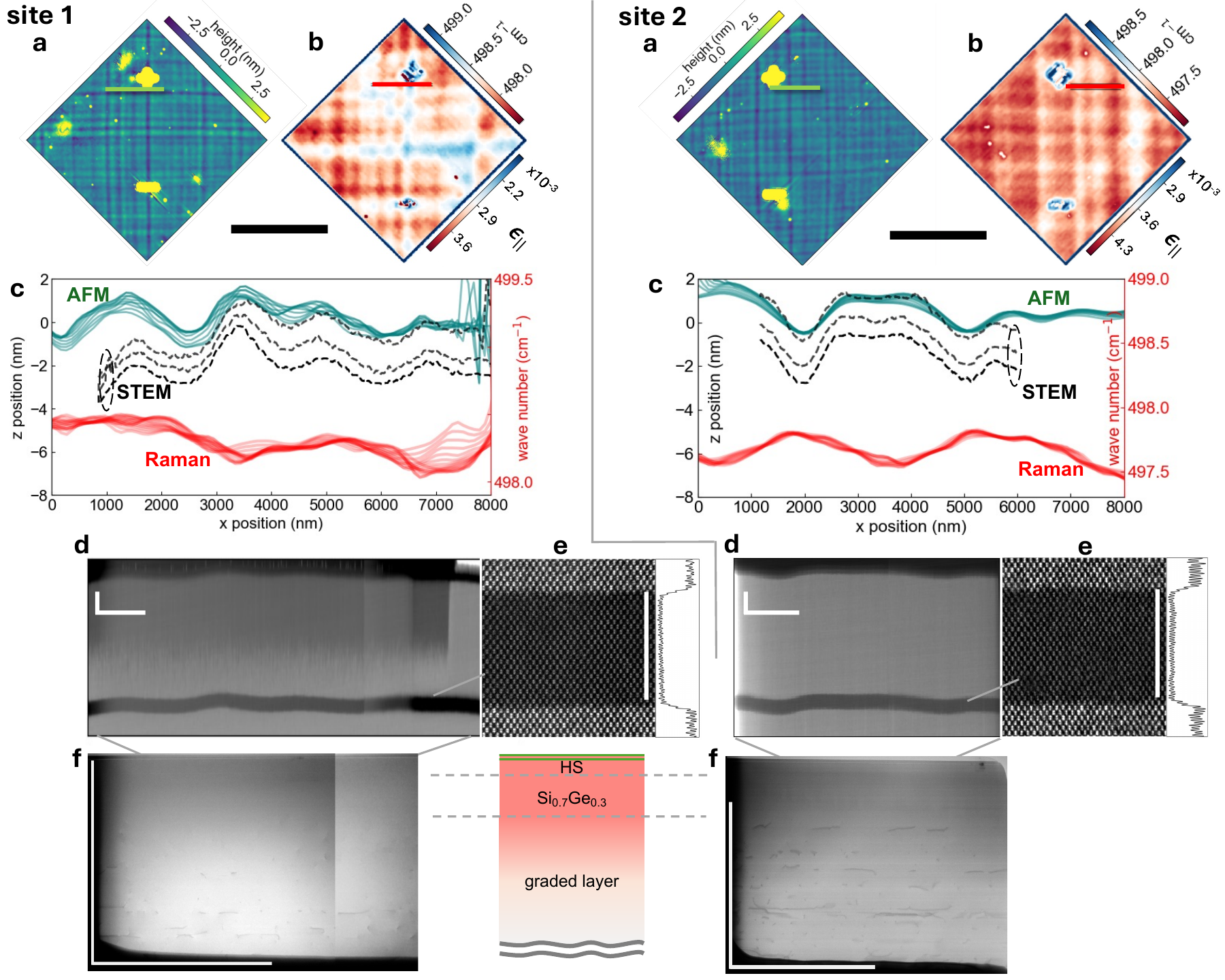}

\caption{\textbf{Cross-correlating heterostructure surface and interface cross-hatch roughness with bulk elastic inhomogeneity using multimodal coincident-site measurements from two sites, 1 (left column) and 2 (right column).} \textbf{a} Heterostructure surface topography images (AFM) measured along with, \textbf{b}, Raman microscopy images, and the corresponding strain, $\epsilon_{||}$. Heavy black scale bars between images are 20 µm and apply to both AFM and Raman data. The colored lines on the AFM and Raman images indicate the locations of cross-sectional lamellae for interface structure analysis (HAADF-STEM). \textbf{c} Plots showing spatially-aligned interface structure, and Raman (strain) measurements at the locations indicated by colored lines in the AFM and Raman images. \textbf{d} Interface nanoscale structure measured using cross-sectional HAADF-STEM imaging. In both cases, the lateral scale bars are 1 µm, and the vertical scale bars are 10 nm. Inset panels \textbf{e} show the atomic-scale interface structure and image intensity line trace across the interface. The 6.0 nm vertical scale bar applies to the image and vertical line trace. \textbf{f} Misfit dislocation bunches, the likely source of the cross-hatch strain fluctuation seen in the Raman data, are distributed randomly throughout the VS SiGe graded layer as shown by coincident-site bright-field TEM images. Horizontal and vertical scale bars are 3~µm  for images from both sites.}

\label{Fig2} 
\end{figure*}

While the cross-hatch topographies shown in Fig.~\ref{Fig1} are similar for all our measured CVD layers, we also observe significant correlations between interfaces within complete heterostructures. Layer deposition remains conformal through more than 50-nm of CVD growth, as evidenced by spatially-aligned structural measurements from multiple perspectives revealing interface-to-interface correlation. Finished heterostructures were imaged with AFM, while buried Si/SiGe interfaces were imaged with cross-sectional HAADF-STEM. Aligned to surface and interface topography, we measured in-plane strain variation $\delta\epsilon_{||}$ using Raman microscopy imaging of Si-Si LO$_z$ peak shifts in SiGe layers. Precision alignment ($\sim$100 nm) between AFM, Raman, and HAADF-STEM images was established using microfabricated metal marks (see Methods and Supplementary Fig. S5). We analyzed a complete heterostructure with a 6.1 nm-thick quantum well grown at 600 \textdegree C. We performed aligned measurements at two sites separated by roughly 100 $\mu$m across the wafer surface. Sites were selected for proximity to metal marks, for alignment accuracy, and presentation of clear topography in AFM images. Figs.~\ref{Fig2} a and b display spatially aligned AFM and Raman images for site 1 (left) and 2 (right). The associated line traces for AFM topography and Raman peak shifts are plotted in Fig.~\ref{Fig2} c, with trace locations indicated in Figs.~\ref{Fig2} a and b. AFM topography and Raman maps show corresponding features. Notably, the deepest trench-like cross-hatch features align with the high Raman Shift values (lower tensile strain on Si-Si bonds, higher compressive strain on SiGe)\cite{10.1063/1.2913052, 10.1063/1.1629142,Kutsukake2004}. Scale bars on Fig. \ref{Fig2} b indicate the strain fluctuation amplitudes, $\delta\epsilon_{||}$.

Regions where cross-sectional lamellae were extracted and imaged by HAADF-STEM coincide with the AFM and Raman line traces in Figs. \ref{Fig2} a and b. Cross-sectional HAADF-STEM images showing the buried s-Si quantum well at sites 1 and 2 are shown in Figs. \ref{Fig2} d, e, and f.  By marking the quantum well interfaces (via an edge-finding algorithm, see Supplementary Fig. S5), the Ångstrom-scale correlation between quantum well interfaces is clear for site 1 and 2,  Figs. \ref{Fig2} c. These data indicate that the 6.1 nm thickness of the quantum well is effectively constant to the Ångstrom-scale over several-$\mu$m-lateral distances, i.e., these two interfaces are highly correlated and the s-Si well is Ångstrom-scale conformal. Any well thickness fluctuation is on the order of single atomic layers as indicated by atomic-resolution images, inset Figs. \ref{Fig2} e, which were measured at a few locations for sites 1 and 2. A line trace to the right of each image shows the mean intensity across the image ($\sim10\times100$~nm$^2$). The rapid oscillations correspond with lateral atomic planes. The interface transition widths of just 3-4 atomic layers (0.4-0.54~nm) indicate near-atomic interface abruptness. Ångstrom-scale interface-to-interface correlation is preserved to the top surface through growth of the 50~nm relaxed SiGe and a 3 nm-thick s-Si capping layer. 

Finally, there is modest anti-correlation between the elastic state and surface/interface undulations (AFM/STEM) revealed by precision-aligned Raman data in Fig. \ref{Fig2}~c. The near-surface strain state is anticipated to vary owing to randomly distributed bunches of misfit dislocations buried in the VS graded layer under the heterostructure, shown in the bright-field TEM images, Figs. \ref{Fig2}~f,  roughly aligned to the other data types from sites 1 and 2.


\subsection{Model: roughening by strain-driven surface diffusion}
The growth study,  Fig. \ref{Fig1} c Wafer \#7 , indicates that surface (interface) roughening evolves overwhelmingly during the pre-growth anneal. 
Moreover, the interface-to-interface correlation in the coincident-site study,  Fig. \ref{Fig2}, indicates that each subsequent heterostructure layer conformally coats roughness on underlying layers. 
We find that a strain-driven surface diffusion model quantitatively predicts the observed roughening dynamics during the anneal, and is also consistent with the interface-to-interface correlation in the heterostructure~\cite{BergamaschiniAPX2016,RovarisPRB2016,RovarisPRB2019,RovarisPRapp2018,Mullins2001}.  

We calculate the dynamical evolution of the free surface of a strained film by a continuum 1 + 1D model, as introduced in Refs.~\cite{RovarisPRB2016,RovarisPRB2019} and schematically depicted in Supplementary Fig. S6.

This approach describes the morphological evolution based on surface diffusion, with the flux of adatoms determined by the gradients in the surface chemical potential $\rho$. The evolution of the surface profile $h$ can be written as:

\begin{equation}
  v_{\perp} = \dfrac{\partial h}{\partial t} = \nabla M\nabla_{\text{s}} \rho
\label{eq:diffusion}
\end{equation}
where $v_{\perp}$ is the perpendicular velocity at each surface position, $M$ is the adatom mobility and the subscript ``s'' denotes the surface gradient operator. The chemical potential $\rho$ in Eq.~\eqref{eq:diffusion} consists of two terms, $\rho = \rho_{\text{s}} + \rho_{\text{el}}$, which are surface free-energy (curvature) and strain-based chemical potentials, respectively.  The derivation of the chemical potentials, and the model details are described in the Methods Section \ref{sec:Morphological_model}.

\begin{figure*}
 \centering
\includegraphics[width=0.9\textwidth]{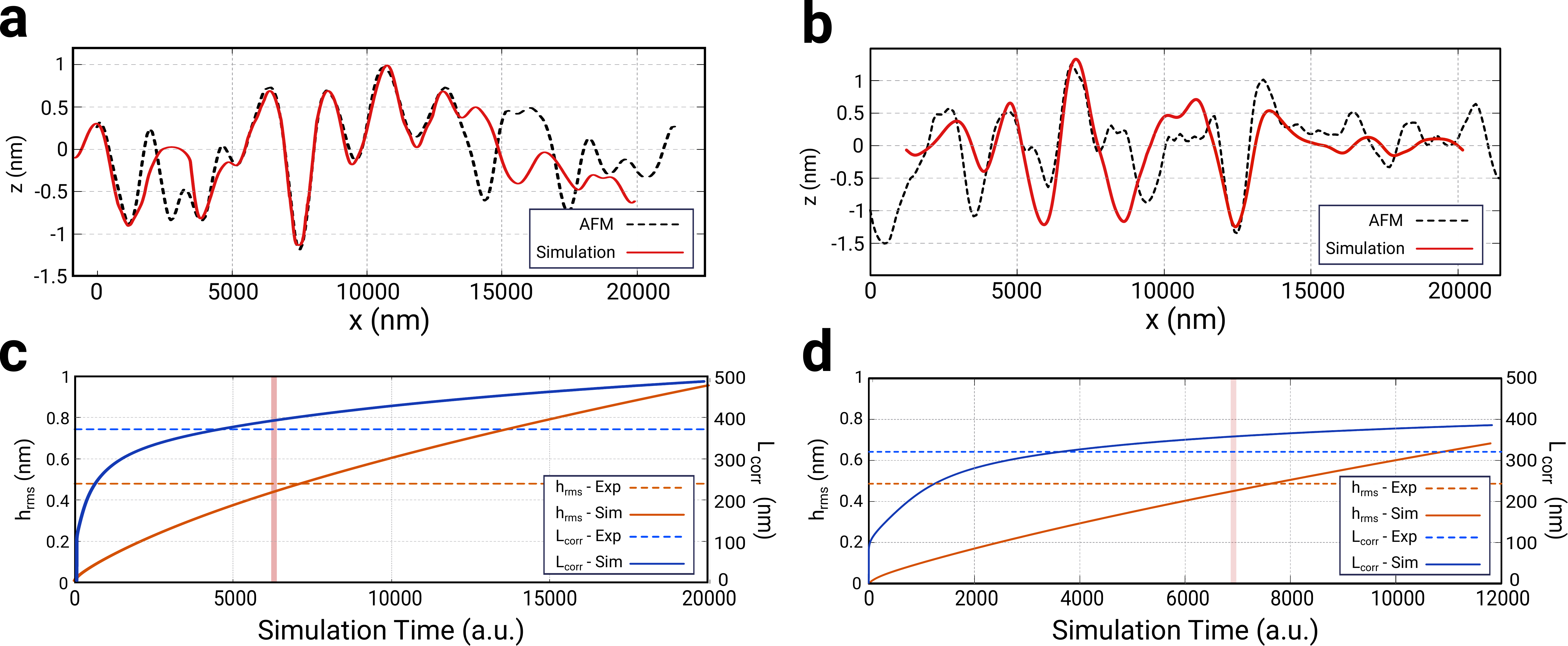}

	\caption{\textbf{AFM-measured surface profiles with simulated counterparts for two selected sites.} Comparison between the simulated profile and the AFM experiments in the first selected region \textbf{a} and in the second selected region \textbf{b}. Comparison between the statistical descriptors of the surface profile, i.e. surface roughness $h_{\text{rms}}$ and correlation length $L_{\text{corr}}$ for the first profile \textbf{c} and the second profile \textbf{d}. The simulation time selected for the best fit with the surface profiles reported in panels \textbf{a} and \textbf{b} is highlighted in red in both plots.}
	\label{Fig3}
\end{figure*}

The model aims at reproducing the surface evolution during the pre-growth anneal step. We model layers with an initially flat profile as a result of the CMP process described in Fig.~\ref{Fig1}. We performed simulated annealing evolutions and compared the results with topography extracted from the AFM data. These latter were recorded on regions of the sample where coincident surface morphology and strain data have been recorded, as shown in Fig~\ref{Fig2} a and c. In Fig.~\ref{Fig3} a and b, the AFM data of the surface profiles for two selected sites are shown in red and are compared with the simulated profiles in black. The criterion used for selecting the simulation snapshot for the comparison was the coincidence of the maximum observed roughness $\Delta h = h_{\text{max}}-h_{\text{min}}$ between the AFM line scans and the simulated profiles. As described in the Methods Section \ref{sec:Morphological_model} we considered different values of average residual strain during our simulations, accounting for a possible non-ideal plastic relaxation of the layer or a thermal strain appearing at the annealing temperature. In both of the cases considered, a small average tensile strain of $\langle \epsilon \rangle$ $=+0.003$ resulted in the best agreement between the simulated profiles and the experimental data. The agreement reported in Fig.~\ref{Fig3} a and b is overall very good, particularly for the comparison of the first site in panel \textbf{a} and generally better in the central region where the influence of the boundary conditions (periodic in the simulations) is less important. 

Furthermore, we confirmed the agreement between simulations and experiments by evaluating statistical descriptors of the surface profiles. The surface root-mean-square ($h_{\text{rms}}$) roughness and the correlation length ($L_{\text{corr}}$) have been evaluated for both the theoretical predictions and AFM data. Results are shown in Fig.~\ref{Fig3} c and d for sites 1 and 2, respectively. The snapshot times selected from the simulations have been highlighted in red. The agreement of the statistical descriptors between theory and experiments is again very good in the comparison made for site 1 shown in Fig.~\ref{Fig3} a, confirming the results observed in the direct profile comparison. The statistical descriptor comparison for site 2 is not so close, especially for the correlation length, but the difference is still limited to about $60~\text{nm}$. We recall that the criterion used for selecting the snapshot has been the closeness to the maximum profile roughness. This is in principle uncorrelated with the surface roughness and the correlation length and thus additionally validates the agreement found between model and experiments.

We explore the model’s validity further by showing that it is consistent with experimental data by predicting the correct approximate timescale to develop surface roughness. In the experiment pre-growth annealing step at 900 \textdegree C, a 1 nm-scale RMS roughness evolves in 120 s. So far, our model has not accounted for a physically-real surface diffusion mobility, (M = 1)  and time (model Simulation Time increments are an arbitrary unit). Using typical physical parameters from independent prior works, reported below, we calculate a surface diffusion mobility, $M$, and driving force, $F=-\triangledown^2 \rho$, for roughening at a rate,  $v_{\perp} = MF$.
The surface mobility $M$ is $M ={V_{a}^2D_{s}c}/{a^2k_BT}$ \cite{Mullins2001}, 
where $V_a$ is the atomic volume, (5.5 \AA)$^3/8$, $D_sc$ is the surface self-diffusion coefficient, $a^2$ is the atomic area, (3.9 \AA)$^2$, $k_B$ is Boltzmann's constant, and $T$ is temperature. At 900 \textdegree C the surface self-diffusion coefficient for Si(100), $D_sc = 1.3\times10^9$ \AA$^2/s$ from which we estimate $M = 4\times10^{11}$ \AA$^6/eVs$ \cite{KEEFFE1994965}. Note that Ge diffusion on Si(100) is estimated to be somewhat faster, $2\times10^9$ \AA$^2$/s, at 900 \textdegree C (extrapolating from data in Ref. \cite{Dolbak2006}). Next, we approximate the driving force factor, $F\sim-\triangledown^2 \rho$. In experiments, the surface is initially flat with a small increasing roughness (curvature) on the timescale of our experiment, and the elastic force dominates, so $F\sim-\triangledown^2 \rho_{el}$. To estimate $\triangledown^2 \rho_{el}$, we model the strain by $\epsilon = \langle\epsilon_\parallel\rangle + \delta\epsilon_\parallel \cos(2\pi x/L)$, where $\langle\epsilon_\parallel\rangle = 0.003$ identified earlier, $\delta\epsilon_\parallel$ is its standard deviation, and L is the correlation length of the strain. From our strain maps, we measure $\delta\epsilon_\parallel = 2\times10^{-4}$ and $L=1.4$ $\mu$m for strain variation along (110). We find $F = -\triangledown^2 \rho_{el} = 8\pi^2K/L^2 [ \delta\epsilon_\parallel^2 \cos(4\pi x/L) + \langle\epsilon_\parallel\rangle \delta\epsilon_\parallel \cos(2\pi x/L)]$, with $K\sim1.6 E$, where $E$ is Young's modulus which is near 120 GPa (0.74 eV/\AA$^2$) for (100) surfaces of Si$_{0.7}$Ge$_{0.3}$ \cite{10.1063/1.1713863}. Combining the driving force and mobility, the peak-to-peak roughness grows at a rate $dh/dt = MF=0.14$ \AA/s, so we forecast $h=1.7$ nm peak-peak height (0.6 nm RMS) after t = 120 s, which is in very good agreement with the experimental RMS roughness values, along with a roughness correlation length of L/4 = 350 nm (determining the distance for diffusion mass transport) that is near experimental values (320 and 370 nm, Fig. \ref{Fig3}).

\subsection{Theory: strain, roughness \& alloy disorder's impacts}

For gate-defined quantum dot spin qubits, our growth study indicates two salient material inhomogeneities:  strain fluctuation of $\sqrt{\langle \epsilon_{\perp}^{2} \rangle} \approx 2\times 10^{-4}$ and nanoscale roughness that is strongly cross-correlated from interface-to-interface. Roughness and strain fluctuations are superposed with atomic-scale alloy disorder, established in prior studies, which we have measured in our heterostructures \cite{Pena2024}. Next, we predict the impacts that all three disorder motifs have on variability across qubit arrays spanning areas larger than micrometers. Strain fluctuation, Figs.~\ref{Fig2}~b and c, changes conduction band offsets between the Si and SiGe layers, affecting the amount of alloy disorder sampled by electrons. A lower band offset allows more of the QD wavefunction to be supported in the SiGe layer, allowing a greater potential effect on valley splitting. Interface roughness adds to the alloy disorder by modifying the alignment of the edge of the Si layer with the valley Bloch functions, affecting valley-orbit coupling and the resulting valley splitting. Moreover, the interface intermixing of Ge atoms into the Si layer also influences valley splitting.

\begin{figure*}
\includegraphics[width=0.9\textwidth]{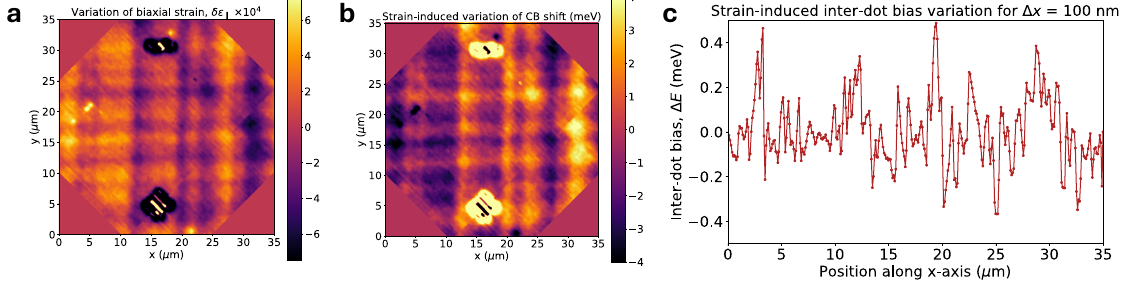}
\caption{\textbf{Strain and conduction band variation inferred from Raman data.} \textbf{a} In-plane biaxial strain $\epsilon_{\parallel}(x,y)$ over the measured sample area, \textbf{b} Conduction band shift $\Delta E_{\mathrm{CB}}(x,y)$ based on deformation theory, \textbf{c} One-dimensional cut through the middle of the measurement region ($y_{0}$=17.5 $\mu$m) of the relative detuning bias shift of the conduction band between points in-plane separated by $\Delta x = 100$ nm along the $x$-axis, $\Delta E_{\mathrm{CB}}(x+\Delta x/2, y_{0}) - \Delta E_{\mathrm{CB}}(x-\Delta x/2, y_{0}) $. This quantifies the inter-quantum dot bias that may be expected due to strain inhomogeneity assuming typical inter-dot separations \cite{Burkard2021}.}
\label{fig:Strain_CB_shift} 
\end{figure*}

Strain effects on quantum dots were modeled in Fig. \ref{fig:Strain_CB_shift} by taking strain data derived from the spatially-resolved Raman mapping and, based on deformation potential theory, estimating the spatially-dependent shift of the conduction band that would be experienced by electrons localized to the s-Si quantum well. To quantify this variability, we considered the effective potential bias shift that would be experienced by a double quantum dot in the well. Our estimates suggest that energy bias variability of roughly tenths of meV may be expected, assuming inter-quantum dot separation of 100 nm.


\begin{figure*}
\includegraphics[width=0.9\textwidth]{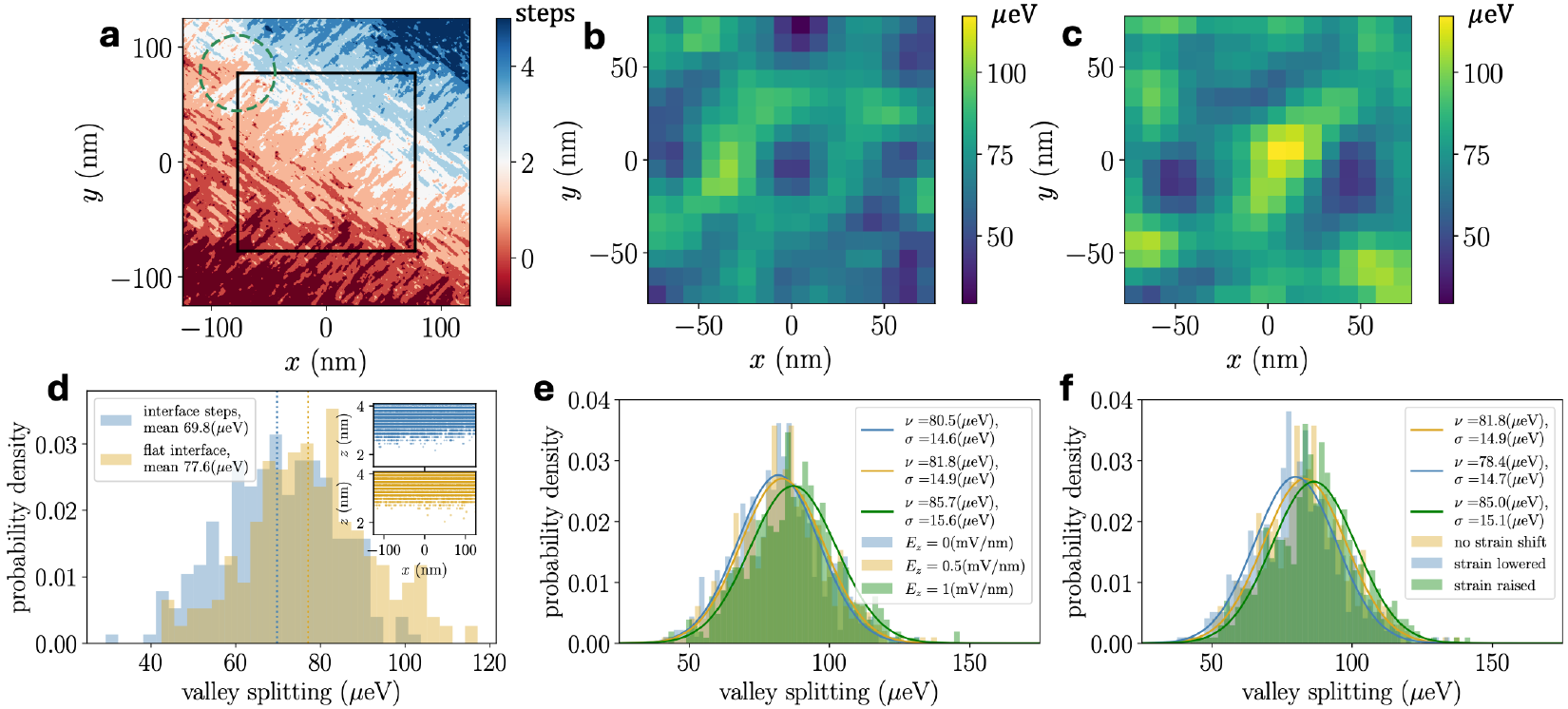}

\caption{\textbf{Valley splitting distributions resulting from interface disorder}
\textbf{a}
STM image of interface steps on the top of the Si layer.
Boxed region used for the calculation in \textbf{b} with dashed green circle of diameter of 65.4 nm, corresponding to the $4\sigma$ width of the ground state of an electron in a harmonic confinement with 1.5 meV spacing.
\textbf{b}
Map of the valley splitting as a function of the center of a QD for a realization of alloy disorder using the interface steps from \textbf{a}.
Note that we consider a conformal step configuration for which both lower and upper interfaces have steps at the same in-plane coordinates.
\textbf{c}
Similar map to \textbf{b}, but with a flat interface between the Si and SiGe layers.
\textbf{d}
Histogram of the valley splittings from \textbf{b} and \textbf{c}.
Calculated valley splitting distribution dependence on \textbf{e} electric field strength and \textbf{f} conduction band offset.
Inset: Sampling of Ge atomic locations around $y=0$ at the Si/SiGe interface for the system with (top) and without (bottom) interface steps.
The conduction band offset is applied to the Si layers to match the extremes of the expected shift from measured strains, and an electric field of 0.5 mV/nm is assumed for all three instances.
Solid curves are fits to Rice distributions, with the associated parameters in the legend.
Histograms of \textbf{e} and \textbf{f} are calculated from 500 alloy realizations, and all valley splitting calculations use a 6.11 nm well thickness, 3.05 nm of the top/bottom SiGe layers included in the computational domain, and assuming a 1.5 meV harmonic in-plane confinement.
}

\label{fig:Valley_splitting} 
\end{figure*}

Next, we calculate how the valley splitting depends on alloy disorder, strain-induced variation of conduction band offset, and electric field strength. 
We set up ensembles of alloy disorder with 6.1 nm well thickness (more specifically, 45 atomic layers), 0.47 nm interface width for the top/bottom of the transition into/out of the SiGe layers (see Supplementary Fig. S7 for derivation of the atomic structure model from image data), 30\% Ge concentration all the way into the SiGe layers, and using a 1.5 meV harmonic in-plane confinement. Fig. \ref{fig:Valley_splitting} e shows histograms of the valley splittings from 500 alloy realizations for 0, 0.5, and 1 (mV/nm) electric fields, as well as the Rice distributions that best fit each histogram and their parameters. To incorporate the effect of strain, Fig. \ref{fig:Valley_splitting} f indicates how those distributions change if one includes the conduction band shifts from Fig. \ref{fig:Strain_CB_shift}, under the assumption that the measured strain is approximately constant across the dot considering the length scales of the measurement. To evaluate the dynamic range one may expect given the strain variation of Fig. \ref{fig:Strain_CB_shift} c, we consider a conduction band shift of $\pm$ 4 meV applied to the Si well. This strain shift has the effect of shifting the band offset relative to the SiGe layers, which may influence the amount of wavefunction penetration into the Si/SiGe barriers.  As shown in Fig. \ref{fig:Valley_splitting} f, we find that this magnitude of strain shift has a negligible effect on the valley splitting distributions.


\begin{figure*}[ht]

\includegraphics[width=0.9\textwidth]{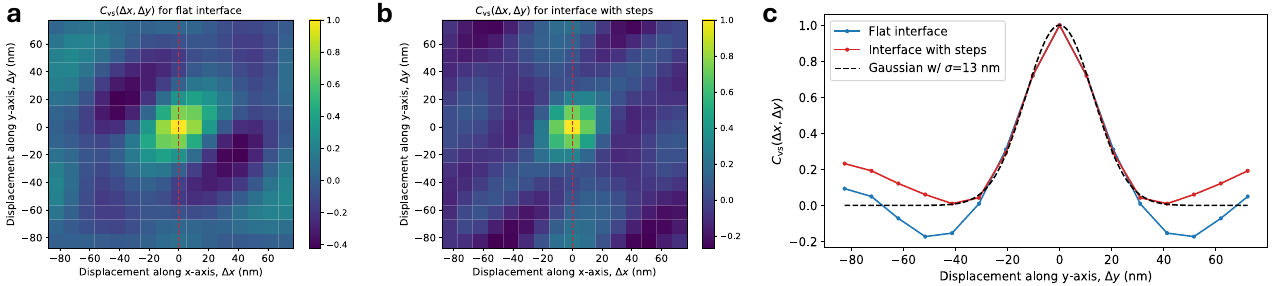}

\caption{\textbf{Spatial correlations of valley splitting} Correlation function of valley splitting for \textbf{a} flat interface and \textbf{b} interface with steps. \textbf{c} One-dimensional cut of $C_{\mathrm{vs}}(\Delta x, \Delta y)$ for $\Delta x=0$, corresponding to the vertical dashed red lines in Figs \ref{fig:VS_correlation_function} \textbf{a} and \textbf{b}. For displacements smaller than the dimensions of the simulation domain, the correlation function decays approximately as a Gaussian with standard deviation 13 nm.}

\label{fig:VS_correlation_function}

\end{figure*}

Lastly, we calculated the degree to which valley splittings in Fig. \ref{fig:Valley_splitting} are spatially correlated as a function of position within the quantum well. To compute the two-dimensional correlation function of valley splitting $C_\mathrm{vs}(\Delta x,\Delta y) = \langle E_{\mathrm{vs}}(x+\Delta x,y+\Delta y) E_{\mathrm{vs}}(x,y)\rangle/\langle E_{\mathrm{vs}}^{2} \rangle$, we make use of the two-dimensional generalization of the convolution theorem. This involves computing the inverse 2D Fourier transform of the square modulus of the 2D Fourier transform of the data. As shown in Fig. \ref{fig:VS_correlation_function}, the correlation function decays with distance in an approximately Gaussian manner for separations that are smaller than the simulation domain. Note that we expect artifacts due to the boundaries to appear for larger spatial separations. As shown in Fig. \ref{fig:VS_correlation_function} c, the correlation of valley splitting decays to good approximation as a Gaussian with standard deviation $\sigma = 13 \ \mathrm{nm}$. This is consistent with the simulated extent of the charge density of the quantum dot wavefunction. Our simulated harmonic confinement gives an orbital splitting of approximately $\Delta = $1.5 meV, which corresponds to a Gaussian wavefunction standard deviation of $\sigma = \hbar / \sqrt{m_{\perp} \Delta} \approx 16 $ nm \cite{Burkard1999}, where $m_\perp = 0.19 m_{0}$ is the transverse effective mass of an electron in Si. The standard deviation of the charge density is lower than that of the wavefunction amplitude by a factor of $1/\sqrt{2}$, corresponding to about 12 nm and consistent with the correlation length we have computed in Fig. \ref{fig:VS_correlation_function}. Our results are consistent with previous experimental and theoretical studies of valley splitting spatial correlations \cite{Losert2023,Volmer2024}.


\section{Discussion}

Across our samples, we find in-plane cross-hatch strain fluctuation,  $\epsilon_{||}=1-1.5\times10^{-3}$ (peak) with correlation length ($\sim1.5\pm0.2$ $\mu$m) comparable to cross-hatch found by Chen ($\sim10^{-3}$),  Sawano ($\sim10^{-3}$), Zoellner (FWHM~$=6\times10^{-4}$), and Esposti ($1.3\times10^{-3}$)~\cite{Chen2002, Sawano2003, Zoellner2015, Esposti2024}. In addition, we find mean tensile strain ($\epsilon_{||}=+0.003$) indicated by the roughening model that agrees reasonably with Raman data (see Supplementary Fig. S3 and S4) ($\epsilon_{||}=0.0015-0.0029$). Zoellner {\it{et al.}} explained that overall tensile strain is likely due to a coefficient of thermal expansion mismatch between the Si wafer and SiGe when cooled from growth temperatures to 300~K~\cite{Zoellner2015}.

Our measured strain fluctuation ($\epsilon_{||}\sim10^{-3}$) is comparable to strain induced by microfabricated MOS gate structures, as reported by Park \textit{et al.} ($\epsilon_{||}\sim0.0003$ (peak) and Corley-Wiciak \textit{et al.} (0.0002-0.0008) using nano-XRD imaging around nano-sized electrodes~\cite{Park2016, Corley-Wiciak2023}. Strain fluctuations are significant disorder sources in quantum devices, e.g., the QuBus which includes many close-spaced electron shuttling gates~\cite{Thorbeck2015,Corley-Wiciak2023}. Cross-hatch strain fields will source a disorder potential landscape sampled locally by each qubit spin. By our estimates, Fig. \ref{fig:Strain_CB_shift} c, strain undulation forces a conduction band fluctuation that is sufficient to create appreciable anomalous variability ($\sim0.1$~meV) in double-dot detunings used to control exchange gates~\cite{Burkhard2023}. Similar effects have been estimated for gate stack-induced strain by Corley-Wiciak \textit{et al.}

Further, cross-hatch strain drives surface roughening effects~\cite{RovarisPRB2016, RovarisPRapp2018, RovarisPRB2019}. Our strain-driven surface diffusion model explains the observed strain-driven cross-hatch morphology at rates consistent with experiments. Providing further validation, the experiments and model indicate that roughness increases overwhelmingly during a higher-temperature annealing (T = 900 \textdegree C) step, whereas lower-T (= 600, 700 \textdegree C) growth (annealing with Si, Ge flux) lead to minimal (\AA ngstrom) changes in roughness and near-perfect interface-to-interface correlation \cite{10.1063/1.1629142, Wong2005}. These observations are explained primarily by a rapid drop in the surface diffusion coefficient, $D_sc = 1.3\times10^9$ \AA$^2/s$, $1.3\times10^7$ \AA$^2/s$, $5.4\times10^5$ \AA$^2/s$ (T = 900, 700, 600 \textdegree C, respectively) and surface mobility, $M$, with thermal activation barrier $2.3\pm0.1$ eV\cite{KEEFFE1994965}. Over the time periods, $\tau$, for annealing and growth (Supplementary Table S2), we estimate surface self-diffusion lengths, $L=\sqrt{D_{s}c  \tau}$, equal to 40~µm (900 \textdegree C) versus 4~µm (700 \textdegree C), and 3~µm (600 \textdegree C), compared to a $\sim2$~µm correlation length for cross-hatch. These values indicate that surface diffusion is amply adequate to explain roughening at 900 \textdegree C/120s but roughly an order-of-magnitude less at 600, 700 \textdegree C even considering the longer growth times. With significantly reduced surface mobility, the response to elastic and surface-curvature driving forces becomes comparatively negligible and each layer evolves as a conformal coating with the close interface-to-interface cross-correlation observed in the cross-sectional experiments, Fig. \ref{Fig2}. Note that more significant deviation from this rule is anticipated for the s-Si well since the mean tensile strain, $\langle\epsilon_\parallel\rangle \sim$ 0.1, is appreciably (3x) larger for this layer and modifies the surface driving force cross-term ($\langle\epsilon_\parallel\rangle \delta\epsilon_\parallel$), but the surface mobility and rapid growth duration (10-100's nm/min) is significantly smaller for T = 600 \textdegree C, and obviates discernible effects as shown by the perfect correlation between interfaces at the top and bottom of the well in the coincident-site study.
Overall, we find that each layer coats conformally, maintaining the epitaxial-layer template that was formed during the 900 \textdegree C anneal, illustrated in Fig. \ref{Fig1}(c). That is, despite the differing layer composition (s-Si vs. relaxed SiGe) and strain state, all layers conformally coat the underlying layers leaving a strongly-correlated roughness that we trace back to the pre-growth anneal.

The cross-hatch roughness creates nonuniform atomic-step distributions at critical interfaces which we predict to impact the valley-orbit physics~\cite{Friesen2010}.
We show that the resulting atomic-step disorder has a mild impact in modifying the valley splitting distribution.
We explore a single example of how interface step disorder changes the valley splitting in Fig.~\ref{fig:Valley_splitting}.
We look at calculations of how these interface steps affect valley splittings with maps of how the splitting varies throughout individual realizations of alloy disorder with and without steps.
The maps show that the average valley splitting differs, with the smooth interface centered around a value of 77 $\mu$eV and the realization with interface steps centered around 70 $\mu$eV.
The slightly lower valley splitting in the calculation with interface roughness is consistent with what would be expected in a material with a longer diffusion length of Ge into the Si layer~\cite{Pena2024}, suggesting that the interface roughness has a similar effect.
Additionally, the correlation length for each map appears to be dictated by the width of the quantum dot wavefunction, as described in the Methods section, 
suggesting that it is the random sampling of the alloy disorder by the wavefunction that determines these correlations \cite{Losert2023,Volmer2024}.
While the results here are compatible with results from previous theory work, we note that an exploration of more alloy disorder realizations is required for more definitive statements.




\section{Conclusion}
Our results show that pregrowth anneals of planarized metamorphic virtual substrates can critically - for quantum applications - roughen the surface despite careful CMP. Our strain-driven surface diffusion roughening model reproduces (1) a reasonable estimate for the details of specific topographic features as well as their statistical descriptors, RMS roughness and correlation length, measured in experiments. (2) Furthermore, using independently-measured surface diffusion data (prior work) the model predicts a rate (time) for 1 nm-sized roughening that is in reasonable agreement with experiment at T = 900 \textdegree C. Finally, negligible roughness evolution occurs in subsequent lower-T (= 600, 700 \textdegree C) growth where significantly lower surface mobility explains the layer-to-layer conformal coating and nearly-perfect interface-to-interface correlations and more abrupt interfaces 3-4 atomic layers (\textit{vs.} 7-10 layers in MBE) observed between each epitaxial layer. Explaining our experimental observations, our model indicates that appreciable roughening occurs overwhelmingly during the pre-growth anneal, while subsequent layers are essentially conformal coatings over a previously roughened surface. The results suggest thicker epitaxial buffer layers underneath heterostructures to isolate the quantum well from the strain inhomogeneity, as well as reduced-temperature (T $\le 700^{\circ}$C) VS processing to inhibit surface roughening during annealing or heterostructure growth.

\section{Methods}

\subsection{SiGe virtual substrate and heterostructure growth}

Strained-Si/Si$_{0.7}$Ge$_{0.3}$ heterostructures on graded virtual substrates were grown at Lawrence Semiconductor Research Laboratory, Inc. via chemical vapor deposition (CVD) using an ASM E2000 Reduced Pressure Reactor and a combination of germane, silane, and dichlorosilane. Virtual substrate Si$_{0.7}$Ge$_{0.3}$ were grown on 100 mm diameter CZ Si $\langle$100$\rangle$ substrates with 10-20 ohm-cm resistivity, P-type (Boron doped), with a linear graded Ge content at $\sim$10\%/\textmu m and a final relaxed Si$_{0.7}$Ge$_{0.3}$ buffer layer 1000 nm thick, growth temperature $\sim$900 \textdegree C. Film stoichiometry was verified by SIMS analysis. Next, substrates are chemically mechanically polished (CMP) to recover surface flatness, which resulted in removal of approximately 500 nm of Si$_{0.7}$Ge$_{0.3}$ buffer layer. A surface cleaning in hydrogen at 900 \textdegree C was carried out to remove residual slurry and adventitious contamination. Heterostructure growth was completed by growing a 225-nm-thick Si$_{0.7}$Ge$_{0.3}$ layer, followed by a 6.1-nm strained Si QW, a 50-nm Si$_{0.7}$Ge$_{0.3}$ spacer layer, and a 3-nm protective Si cap. Heterostructure layers were grown at 600 \textdegree C, with growth rates of 7 nm/min for SiGe and 0.2 nm/min for Si. STM imaging was performed with a {\it Scienta Omicron Lab10 MBE, VT STM} system (tunnel current I$= 0.2-0.5$nA, tip bias $-2.0$ to $-2.5$V) at a few sites on the sample. Additional details on AFM and Raman imaging are in the Methods and Supplementary Figs. S1,  S3, and S4.

\subsection{Coincident site measurements}
The AFM, HAADF-STEM, and Raman image data are precision-aligned to metal fiducial markers microfabricated using a photolithography and liftoff process. Photolithography was performed with laser-based (Heidelberg ML6) process. Next, metal (5 nm-thick Ti then 50 nm-thick Au) was deposited by electron-beam evaporator. The markers were defined by metal liftoff. Prior to AFM measurements, the sample was cleaned in acetone and isopropyl alcohol and blasted dry with N$_2$. Details of the AFM, Raman, and HAADF-STEM imaging are described in the Supplementary Methods. In brief:

The AFM (Digital Instruments DI 3100) imaging was performed in air with µmasch NSC-15 probes. Image background artifacts were corrected by line-by-line subtraction of a fourth-degree polynomial.

We performed Raman imaging tracking the Si-Si LO$_z$ peak in the SiGe layer, which is a sensitive probe for both composition and strain~\cite{DeWolf1996, Mermoux2010}. Measuring Raman with polarized light in a backscatter configuration at normal incidence allows direct sensitivity to in-plane strains, $\epsilon_{||}$ owing to selection rules~\cite{DeWolf1996}. Predominant in-plane biaxial strains, due to stresses from the underlying layers, are the most significant macroscopic elastic deformations anticipated in flat, planar, epitaxy. Raman imaging was performed using a WITec Alpha300R with a 488 nm laser and a 100X/0.95 NA objective. Each sample was measured a total of 9 times by examining 3 separate locations. For each measurement, a $25\times25$ µm$^2$ region was examined with a spectrum collected every 333 nm. For each sample, locations near each edge and center were acquired. 
Spectral accuracy is $\pm0.17$ cm$^{-1}$ and precision is $<0.05$ cm$^{-1}$. Each spectrum was fit with a Voigt function near the Si-Si LO$_z$ peak $\sim500$ cm$^{-1}$ by fitting only in this region. Raman microscopy images are plots of the fitted Si-Si peak position in the SiGe layer. We estimate that the 488 nm light returns appreciable signal up from depths up to roughly 200 nm. The absorption coefficient for 488 nm (2.54 eV) light in Si$_{0.7}$Ge$_{0.3}$ is $\alpha\simeq5.5\times 10^4$~/cm~\cite{Jellison1993}, and a typical rule of thumb is to take a probe depth where $90\%$ of the Raman signal originates, which is $1.15/\alpha\simeq 200$~nm~\cite{DeWolf1996}. 

Prior works show the Si-Si LO$_z$ peak, $\omega$, in the Si$_{1-x}$Ge$_{x}$ alloy shifts with composition, $x$, and strain, $\epsilon_{||}$, by $\omega = 520.7 - 66.9x - 730(70)\epsilon_{||}$ with comparable relations reported by several other works.\cite{10.1063/1.2913052, Tsang1994, Perova2011, Rouchon2014} Using $x=0.31\pm0.01$ from our SIMS data (Supplemenentary Fig. S2),  yields mean tensile strains $\sim 0-0.005$  with  cross-hatch strain RMS fluctuations $=0.00032(4)$.





Lamellae for HAADF-STEM were prepared using ion milling and lift-out with a {\it ThermoFisher Scientific Helios Nanolab 660} dual-beam focused ion beam (FIB) and final thinning performed at 1 keV using Ga as the milling species. HAADF-STEM images were acquired with a {\it Hitachi HD2700} probe-corrected STEM using an electron beam energy of 200 keV with detector inner and outer angles of 65 mrad and 271 mrad respectively. 

%

\subsection{Model: morphological evolution of strained films \label{sec:Morphological_model}}

%
%
The chemical potential $\rho$ in Eq.~\eqref{eq:diffusion} consists of two terms, $\rho = \rho_{\text{s}} + \rho_{\text{el}}$. The first one, $\rho_{\text{s}}$, represents the energy cost associated with the formation of new free surfaces and, for isotropic surface energy density $\gamma$, is linearly proportional to the local surface curvature $\kappa$, $\rho_s \approx \kappa\gamma$. In this paper we considered $\gamma_{\text{Ge}} = 60~\text{meV/\AA}$ for Ge(001) and $\gamma_{\text{Si}} = 87~\text{meV/\AA}$ for Si(001). Linear interpolation was considered for evaluating the surface energy density of the Si$_{0.7}$Ge$_{0.3}$ alloy. 

The second term, $\rho_{\text{el}}$, is proportional to the elastic energy density. The strain value at the free surface of a plastically-relaxed thin film is generally evaluated by summing all the strain fields due to the array of dislocations relaxing the layer~\cite{RovarisPRapp2018} and by solving the partial differential equation for the mechanical equilibrium of the thin film. In the context of this work, the strain data are directly derived from spatially-resolved experimental Raman measurements. These data record the in-plane strain $\epsilon_{\parallel}$ in the layer and have been used to compute the surface chemical potential. We exploited two assumptions to determine the full strain tensor: (i) the two in-plane components of the strain, $\epsilon_{xx}$ and $\epsilon_{zz}$, have the same average value so that:
\begin{equation}
    \epsilon_{\parallel} = \dfrac{\epsilon_{xx}+\epsilon_{zz}}{2} \qquad \text{with} \quad \langle\epsilon_{xx}\rangle\approx\langle\epsilon_{zz}\rangle 
\label{eq:parallel}
\end{equation} 
(ii) the out-of-plane component has been determined as for a flat film since the typical value of the surface corrugation $h_{\text{rms}}$ is much lower than their typical wavelength $L$,  $h_{\text{rms}}/L \ll 1$. 

Given this spatially-varying and isotropic in-plane strain $\epsilon_\parallel$, we infer the resulting out-of-plane strain $\epsilon_{\perp} = \epsilon_{zz}$ by assuming zero imposed out-of-plane stress $\sigma_{zz} = 0$. From Hooke's law $\sigma = C \epsilon$, we obtain
\begin{equation}
	\epsilon_{\perp} = \dfrac {-2c_{12}}{c_{11}} \epsilon_{\parallel} \approx -0.772 \epsilon_{\parallel},
    \label{eq:perp}
\end{equation}
where the elastic moduli of silicon are $c_{11} = 165.6 \ \mathrm{GPa}$ and $c_{12} = 63.9 \ \mathrm{GPa}$.\cite{Hall1967}
%

Finally, we used a polynomial interpolation to smoothly connect the strain data at the boundaries of the measured region, consistently with the periodic boundary conditions applied at model boundaries (like shown in Supplementary Fig. S6 a).

With the above-described assumptions, the elastic chemical potential can be evaluated from the measured strain tensor $\epsilon$ as:
\begin{equation}
    \rho_{\text{el}} = V_a\cdot\mu\Big{(}\sum_i\epsilon_{ii}^2+\sum_{i\neq j}\epsilon_{ij}^2\Big{)}+\lambda/2\Big{(}\sum_i \epsilon_{ii}\Big{)}^2
\end{equation}
where $\mu$ and $\lambda$ are the Lamé constant. 

The chemical potential is then plugged into Eq.~\eqref{eq:diffusion}, and its solution is used to reproduce the dynamics of the surface evolution by surface diffusion. In this work, we  also considered the effect of an average residual strain by adding a constant term to the strain tensor evaluated as described in equations~\eqref{eq:parallel} and~\eqref{eq:perp}. We considered different values of average strain in the window $[-0.3\%:0.3\%]$ to take into consideration the existence of a residual unrelaxed strain appearing at the annealing condition, due to non-ideal plastic relaxation or thermal strain. 

We started from flat profiles and performed simulated annealing evolutions until the surface roughness value corresponded to the experimental measures. The typical evolution of the maximum surface roughness (calculated as $\Delta h = h_{\text{max}}-h_{\text{min}}$ is reported in Supplementary Fig. S6 b for three different values of the average in-plane strain. As can be seen in Supplementary Fig. S6 b, the annealing simulations show an increase in the surface roughness value with respect to the annealing time. Significant snapshots extracted from these simulations are shown in Supplementary Fig. S6 c-d for the cases $\epsilon_{\parallel} = 0 \%$ and $\epsilon_{\parallel} = +0.3\%$, respectively. As can be appreciated in the two panels, the surface dynamics and the typical features produced by the annealing are quite different in the two cases, highlighting the influence of the average residual strain.   

\subsection{Model: strain disorder potentials }

To map from Raman shift to strain, we assume an average Raman shift coefficient {\it{b}}$^{\textrm{Si-Si}}$ due to in-plane strain \cite{10.1063/1.2913052}, where

\begin{equation}
	\epsilon_{\parallel} \approx -\dfrac {(\omega-\overline{\omega})} {730~{\text{cm}}^{-1}}.
\end{equation}

From these axial strain components we infer the resulting spatial variation of the conduction band by appealing to deformation potential theory,
\begin{eqnarray}
	\Delta E_{\pm z} & = & \Xi_d \epsilon_{xx} + \Xi_d \epsilon_{yy} + (\Xi_d + \Xi_u)\epsilon_{zz} \\
	& = & (2\Xi_d)\epsilon_{\parallel} + (\Xi_d + \Xi_u)\epsilon_{\perp},
\end{eqnarray}
where $\Xi_d \approx$ 1.1 eV and $\Xi_u \approx$ 10.5 eV are the dilatation and shear deformation potentials, respectively \cite{Fischetti1996}.

\subsection{Model description: Valley splitting ensemble calculations}

The valley splitting ensembles of Fig. \ref{fig:Valley_splitting} were calculated using an envelope function framework where valley splitting is assumed to be caused by the localized repulsion of Ge atoms.
The wavefunction of an electron in our quantum dot can be written as
\begin{equation}
\psi\left(\pmb{r}\right)=\sum_ve^{i\pmb{k}_v\cdot\pmb{r}}u_v\left(\pmb{r}\right)F_v\left(\pmb{r}\right),
\end{equation}
where $v$ indexes valleys, $\pmb{k}_v$ is the wavevector associated with valley $v$, $u_v$ is the corresponding Bloch function, and $F_v$ is an function that describes the envelope of the wavefunction in its associated valley.
This can be used to determine a Hamiltonian acting on the envelope functions for an electron in the $\pm z$ valleys of Si, and the block of this Hamiltonian coupling valleys $v_1,v_2\in\{+z,-z\}$ can be expressed as

\begin{align}
    \hat{H}_{v_1v_2}=&\delta_{v_1v_2}\left(\frac{1}{2}\hat{\pmb{p}}\cdot\hat{m}_{v_1}^{-1}\hat{\pmb{p}}+V_\mathrm{ext}\right)\\&+\alpha\sum_je^{i\left(\pmb{k}_{v_1}-\pmb{k}_{v_2}\right)\cdot\pmb{r}}u_{v_1}^*\left(\pmb{r}\right)u_{v_2}\left(\pmb{r}\right)\delta^{(3)}\left(\pmb{r}_j-\pmb{r}\right),
\end{align}
where $\hat{\pmb{p}}$ is a vector of momentum operators, $\hat{m}_ v$ is the effective mass matrix for electrons in valley $v$, $V_\mathrm{ext}$ is the electric potential applied by nearby gates, and $\pmb{r}_j$ is the location of a Ge atom.
We set the repulsive strength of individual Ge atoms as $\alpha=$12(meV$\cdot$nm${}^3$) to match the band offset of 600(meV) between pure Si and pure Ge.
Alloy realizations are sampled by randomly selecting whether each atomic site in the computational domain is a Si or Ge atom based on the spatially dependent Ge concentration and then used to calculate the energy levels for the two lowest states for the valley splitting.


\section{Acknowledgement}

We would like to thank R. Butera and C. Richardson of University of Maryland, C. Carter, Emeritus Professor U. of Connecticut, and T. Lu at Sandia National Laboratories for thought-provoking discussion and critical reading of the manuscript.
This work was performed at the Center for Integrated Nanotechnologies, an Office of Science User Facility operated for the U.S. Department of Energy (DOE) Office of Science. 
Research supported as part of $\mu$-ATOMS, an Energy Frontier Research Center funded by the U.S. Department of Energy (DOE), Office of Science, Basic Energy Sciences (BES), under award DE-SC0023412 (data analysis and manuscript preparation). Sandia National Laboratories is a multi-mission laboratory managed and operated by National Technology and Engineering Solutions of Sandia, LLC, a wholly-owned subsidiary of Honeywell International, Inc., for the U.S. DOE’s National Nuclear Security Administration under contract DE-NA-0003525. This paper describes objective technical results and analysis. Any subjective views or opinions that might be expressed in the paper do not necessarily represent the views of the U.S. Department of Energy or the United States Government.

\section{Competing Interests}
The authors declare no competing interests.\\

\section{Author Contributions}

L. Peña: investigation, formal analysis, writing - original draft;
M. Brickson: theoretical analysis, electronic structure calculations, writing;
F. Rovaris: theoretical analysis, writing - original draft;
J. Houston Dycus: investigation, formal analysis, writing - review and editing; 
A. McDonald:investigation, formal analysis, writing - review and editing; 
Z. T. Piontkowski: investigation, formal analysis, writing - review and editing; 
J. Ruzindana: investigation;
A. Bradicich: formal analysis, writing - review and editing;
D. Bethke: investigation;
R. Scott: investigation, formal analysis, writing - review and editing; 
T. Beechem: investigation, formal analysis, writing - review and editing; 
F. Montalenti: theoretical analysis, writing - review and editing;
N. Jacobson: conceptualization, theoretical analysis, software, writing - original draft, funding acquisition, project administration; 
E. Bussmann: conceptualization, formal analysis, writing - original draft, funding acquisition, project administration.

\bibliography{bibliography.bib}

\end{document}


\title{Supplementary information: Cross-hatch strain effects on SiGe quantum dots for qubit variability estimation} 

\author{Luis Fabián Peña}
\affiliation{Sandia National Laboratories, Albuquerque NM, USA}
\altaffiliation{Department of Physics, Baylor University, Waco, TX 76798, USA}

\author{Mitchell I. Brickson}
\affiliation{\mbox{Center for Computing Research, Sandia National Laboratories, Albuquerque NM, USA}}

\author{Fabrizio Rovaris}
\affiliation{Department of Materials Science, University of Milano-Bicocca, Milano, Italy}

\author{J. Houston Dycus}
\affiliation{Advanced Microscopy, Eurofins EAG Materials Science, Raleigh NC, USA}

\author{Anthony McDonald}
\affiliation{Sandia National Laboratories, Albuquerque NM, USA}

\author{Zachary T. Piontkowski}
\affiliation{Sandia National Laboratories, Albuquerque NM, USA}

\author{Joel Benjamin Ruzindana}
\affiliation{\mbox{Department of Chemistry and Physics, University of Arkansas at Pine Bluff, Pine Bluff AR, USA}}

\author{Adelaide M. Bradicich}
\affiliation{\mbox{Center for Integrated Nanotechnologies, Sandia National Laboratories, Albuquerque NM, USA}}

\author{Don Bethke}
\affiliation{\mbox{Center for Integrated Nanotechnologies, Sandia National Laboratories, Albuquerque NM, USA}}

\author{Robin Scott}
\affiliation{\mbox{Lawrence Semiconductor Research Laboratory, Inc., Tempe AZ, USA}}

\author{Thomas E. Beechem}
\affiliation{\mbox{Mechanical Engineering and Birck Nanotechnology Center, Purdue University, West Lafayette IN, USA}}

\author{Francesco Montalenti}
\affiliation{Department of Materials Science, University of Milano-Bicocca, Milano, Italy}

\author{N. Tobias Jacobson}
\email{ntjacob@sandia.gov}
\affiliation{\mbox{Center for Computing Research, Sandia National Laboratories, Albuquerque NM, USA}}

\author{Ezra Bussmann}
\email{ebussma@sandia.gov}
\affiliation{Sandia National Laboratories, Albuquerque NM, USA}

\date{\today}

\keywords{SiGe, quantum dot, qubit, strain, CVD, AFM, HAADF-STEM}




\maketitle{}
\section{Supplementary methods - Growth study}
\beginsupplement

The growth study tracked surface evolution of 25 wafers undergoing a standard commercial chemical vapor deposition process, with details covered in the Methods. Table I indicates the layer stacks and growth temperatures prepared for this study. Table II indicates the growth rates across the wafer set.

\begin{table*}[h!]
  \caption{Growth study tracking growth surface morphology and strain evolution of 25 wafers through steps in a standard commercial CVD process}
  \label{tbl:1}
\resizebox{\textwidth}{!}{
  \begin{tabular}{llll}
    \hline
    \hline
    Wafer     &~~Process																									&Characterization    	&\\ 	
    \hline
    1, 2          &Si(100) substrate		       																				&~~STM				&\\			
    3, 4          & Graded layer + relaxed buffer																			&~~AFM				&\\	
    5, 6          & CMP of virtual substrate																					&~~AFM				&\\	
    7             & LSRL Prep																									&~~AFM				&\\ 		
    8             & T = 600 $^{\circ}$C, SiGe regrowth 5 nm																&~~AFM				&\\ 
    9             & T = 600 $^{\circ}$C, SiGe regrowth 70 nm															&~~AFM				&\\ 
   10             & T = 600 $^{\circ}$C, SiGe regrowth 70 nm/ well 10 nm												&~~AFM				&\\ 
   11             & T = 600 $^{\circ}$C, SiGe regrowth 70 nm/ well 10 nm/ SiGe 50 nm/ Si cap 3 nm				&~~AFM				&\\ 
   12            & T = 600 $^{\circ}$C, SiGe regrowth 225 nm/ well 3 nm												&~~AFM				&\\ 
   13            & T = 600 $^{\circ}$C, SiGe regrowth 225 nm															&~~AFM				&\\ 
   14            & T = 600 $^{\circ}$C, SiGe regrowth 225 nm/well 10 nm												&~~AFM				&\\ 
   15            & T = 600 $^{\circ}$C, SiGe regrowth 225 nm/well 10 nm/ SiGe 50 nm/Si cap 3 nm				&~~AFM, STEM, Raman&\\ 
   16            & T = 600 $^{\circ}$C, SiGe regrowth 225 nm/well 15nm												&~~AFM				&\\ 
   17             & T = 600 $^{\circ}$C, SiGe regrowth 500 nm															&~~AFM				&\\ 	
   18             & T = 700 $^{\circ}$C, SiGe regrowth 5 nm																&~~AFM				&\\ 
   19             & T = 700 $^{\circ}$C, SiGe regrowth 70 nm															&~~AFM				&\\ 
   20            & T = 700 $^{\circ}$C, SiGe regrowth 500 nm															&~~AFM				&\\ 	
   21            & T = 700 $^{\circ}$C, SiGe regrowth 225 nm/well 3 nm												&~~AFM				&\\ 
   22             & T = 700 $^{\circ}$C, SiGe regrowth 225 nm/well 15 nm												&~~AFM				&\\ 
   23             & T = 700 $^{\circ}$C, SiGe regrowth 225 nm															&~~AFM				&\\ 
   24             & T = 700 $^{\circ}$C, SiGe regrowth 225 nm/ well 10 nm											&~~AFM				&\\ 
   25             & T = 700 $^{\circ}$C, SiGe regrowth 225 nm/ well 10 nm/SiGe 50 nm/Si cap 3 nm				&~~AFM				&\\ 
\hline  
\end{tabular}}
Note: The actual Si QW thickness for Wafer \#15 is 6.1 nm, as measured with TEM
\end{table*}

\begin{figure*}
\includegraphics[width=5.5 in]{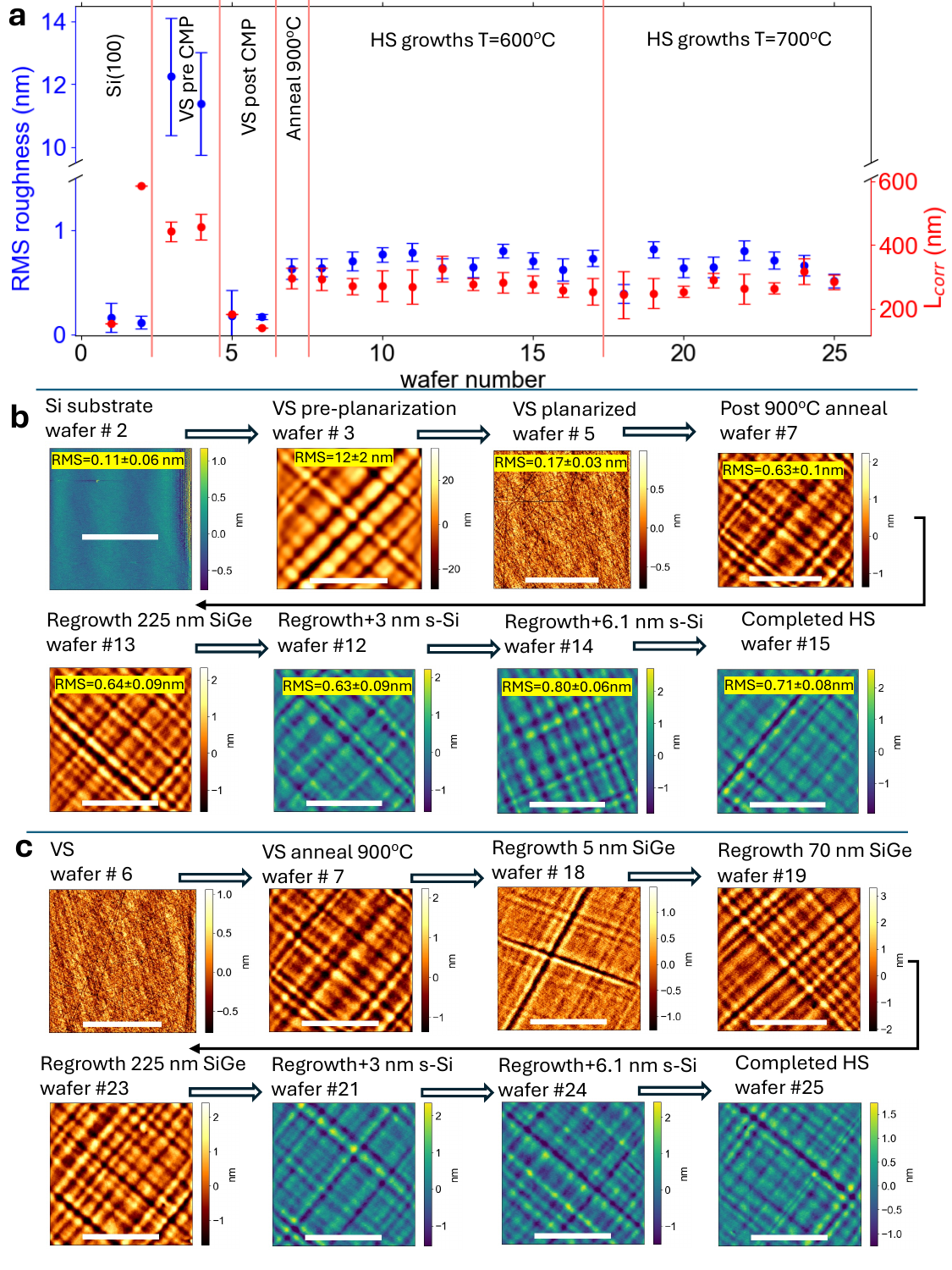}
\caption{
\textbf{Tracking growth surface roughness for the 25 wafer CVD process,} through virtual substrate (VS) preparation, the pre-growth anneal, $T = 900^{\circ}$C, and heterostructure (HS) depositions at 600 $^{\circ}$C and 700 $^{\circ}$C.  \textbf{ a} The calculated root-mean-square (RMS) roughness, and correlation length, $L_{corr}$, for each wafer with one standard deviation, $\sigma$, error bar (among image rows). The $L_{corr}$ error bar is suppressed for clarity for wafers \#1, 2, 5, and 6, because it is appreciable owing to difficulty measuring uncorrelated structure of near-atomically flat surfaces. \textbf{ b} Growth surface evolution at $T=600^\circ$C, tracked by AFM imaging after each step. left to right from upper row: Si(100) substrate, relaxed Si$_{0.7}$Ge$_{0.3}$ on graded growth (VS pre CMP), and the VS post CMP surface, and the planarized VS after $T=900^\circ$C pre-growth anneal. Bottom row, left to right: HS growth on epitaxy-ready Si$_{0.7}$Ge$_{0.3}$ VS following regrowth of 225 nm SiGe, then the 3 nm and 6.1 nm s-Si well, and finally the completed HS, after capping the well with 50 nm thick SiGe and 3 nm s-Si. All scale bars 10 $\mu$m. \textbf{c} Growth surface evolution at 700 $^{\circ}$C. These images show that starting from the initially flat VS post CMP (wafers \#5 and \#6), the pregrowth anneal roughens the surface (RMS~$=0.63\pm0.1$~nm) with cross-hatch features, then all subsequent growth steps have comparable crosshatch roughness. That is, roughness is predominately introduced at the anneal (wafer \#7), then all subsequent growth process steps do not distinguishably change the roughness and correlation length (wafers \#8-25). The roughness stays near $0.6$~nm, with three outliers at $0.4$~nm and $0.8$~nm,  while  $L_{corr}\sim300$~nm. Horizontal scale bar is 10 µm.
}
\label{SI FigS1}
\end{figure*} 

\begin{table}[h!]
\centering
\caption{Growth temperatures and rates for the various wafers and layers.}
\label{tab:growth_conditions}
\begin{tabular}{lcc}
\toprule
Layer & growth $T$ (\textdegree C) & Growth rate (nm/min) \\
\hline

\multicolumn{3}{l}{Wafers 8-17} \\
\quad SiGe & 600 & 7 \\
\quad Si & 600 & 0.2 \\

\multicolumn{3}{l}{Wafers 18-25} \\
\quad SiGe & 700 & 75 \\
\quad Si & 700 & 4 \\
\hline

\hline
\end{tabular}
\end{table}

\begin{figure}[h!]
\includegraphics[width=3.9 in]{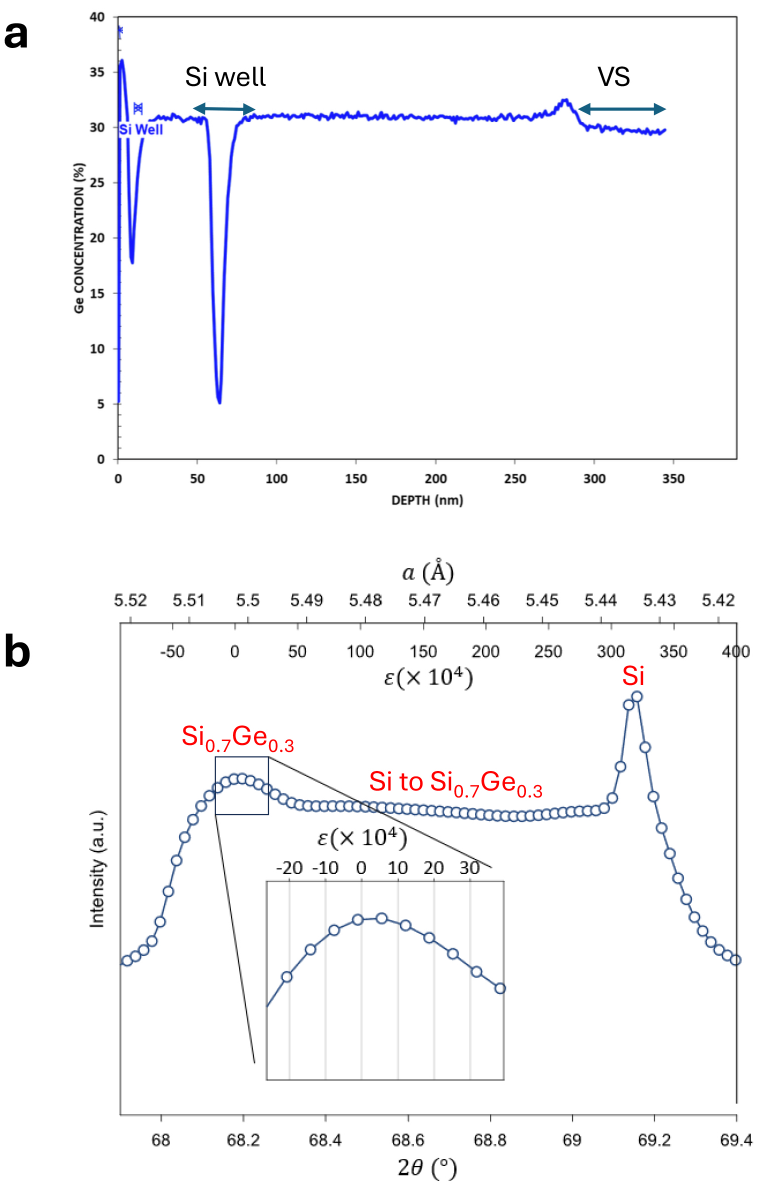}
\caption{\textbf{Measurements indicating mean composition and elastic state of the of the heterostructure in this study.} \textbf{a} Secondary ion mass spectroscopy (SIMS) depth profile indicates the alloy Ge$_x$ content to be $x\sim0.31$ with a buried Si well at the expected depth $\sim55$ nm below the surface, on a $\sim225$ nm-thick SiGe buffer atop the virtual substrate (VS). The heterostructure and VS compositions are similar. \textbf{b} An X-ray diffraction (XRD) $\omega - 2\theta$ scan (rocking curve) of the heterostructure shows an on average relaxed SiGe thin film on the graded SiGe buffer and Si substrate. The scan was performed on a Rigaku Smartlab X-Ray Diffractometer, and the strain states indicated in the top (below) x-axis were simulated (Rigaku GlobalFit software) for strained and strain-relaxed Si$_{0.7}$Ge$_{0.3}$. Lattice constants corresponding to the $2\theta$ position are indicated in the top (above) x-axis for reference.}
\label{SI Fig2} 
\end{figure}

\subsection{Raman imaging}


To quantify near-surface strain fluctuations, we performed a strain mapping using Raman microscopy following studies that established a formalism to use Raman modes (Si-Si LO$_z$ shift in the SiGe) to measure the in-plane strain state, $\epsilon_{||}$, in the epitaxial SiGe \cite{10.1063/1.2913052,Baribeau1995, SHIN2000505}. The essential experimental details are described in the Methods section. Here we explain the analysis in more detail.

\begin{figure*}[h!!]
\centering
\includegraphics[width=5 in]{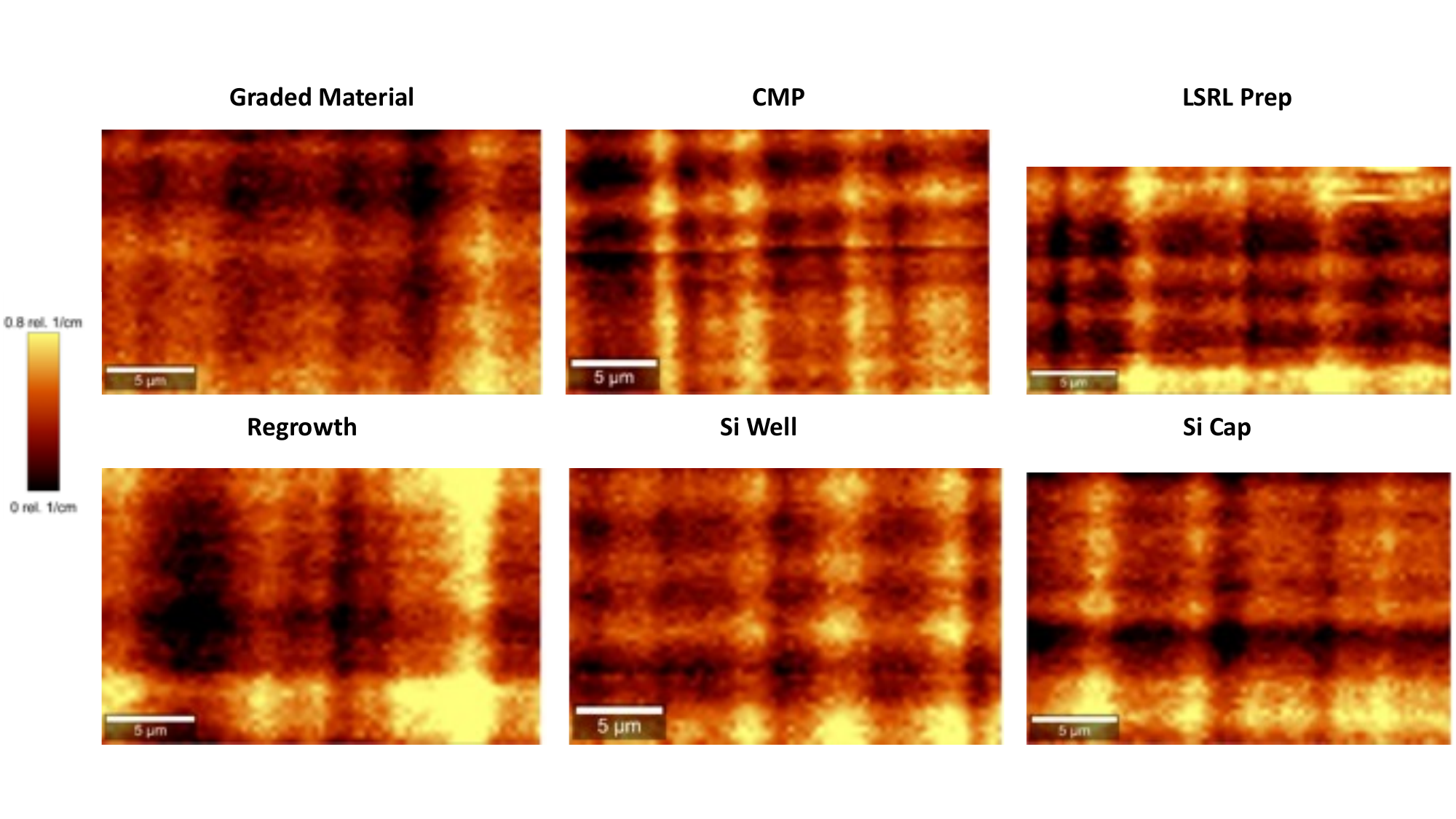}
\caption{\textbf{To assess near-surface strain variations, Raman microscopy was employed following established formalisms that relate Si–Si vibrational mode shifts in SiGe to the in-plane strain state within the top ~200 nm.} Raman images of Si-Si peak position at each process step compared to AFM topographies measured in Fig. \ref{SI Fig4}.  Magnitude of range for each image is equivalent (0.8 cm$^{-1}$).  }
\label{SI Fig3}
\end{figure*}

\begin{figure*}
\includegraphics[width=6 in]{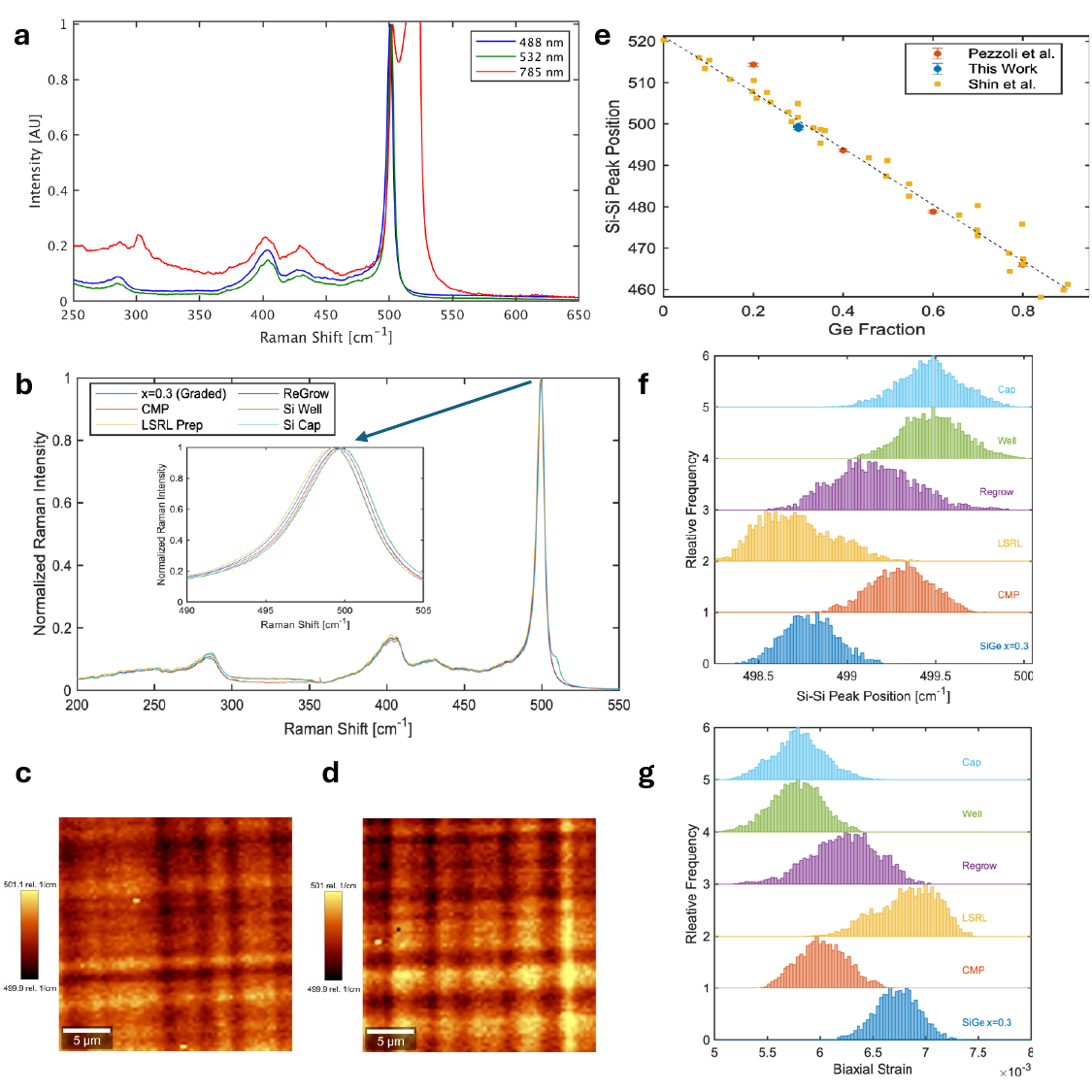}
\caption{\textbf{Raman-based strain analysis throughout the growth process, correlating Si–Si peak shifts with AFM topographies and highlighting spectral evolution across different stages.} Raman strain analysis of Si-Si peak position to AFM topographies measured in SI Fig. \ref{SI Fig4}. \textbf{a} Raman spectra obtained at various wavelengths from a completed polished virtual substrate. \textbf{b} Spectra characterizing samples following each stage of the growth process (T = 600 $^{\circ}$C). Each curve is the mean normalized spectra across ~5625 spectra making up each strain-mapping image. Inset: Zoom-in of Si-Si vibration of SiGe near 500 cm$^{-1}$ where distinct shifting is observed at each step of the process.  These shifts are much greater than experimental uncertainties. \textbf{c-d} Two  examples plots (‘Raman’ images) of the Si-Si mode’s peak position taken with \textbf{c} 488 nm and \textbf{d} 532 nm laser wavelengths. \textbf{e} Our result for Si-Si Raman peak shift plotted alongside data from Shin and Pezzoli \cite{10.1063/1.2913052, SHIN2000505}. \textbf{f} Histograms of Si-Si Raman peak positions over entirety of each dataset for each stage of the growth process. \textbf{g} Histograms of strain fluctuations calculated from Si-Si  Raman peak positions.}
\label{SI Fig4}
\end{figure*}

Raman measurements reported in the manuscript were obtained using 488 nm light (2.54 eV). Raman strain mapping was performed for several wafers, Fig. \ref{SI Fig3}.   Initially, we tried excitation wavelengths 488, 532, and 785 nm. Typical Raman spectra are indicated in Fig \ref{SI Fig4} a. The 785 nm light probed both the heterostructure and the Si substrate, as indicate by the double peaks around 500 cm$^{-1}$ and 520 cm$^{-1}$. By contrast, both the 488 nm and 532 nm excitation yield well-defined single-peaked Raman responses consisted with prior studies on SiGe \cite{10.1063/1.2913052,Baribeau1995, SHIN2000505}. 


Fig. \ref{SI Fig4} b shows the average of 5,265 spectra for each sample, characterizing the mean strain state after each process step. The inset highlights a robust and measurable peak shift, significantly exceeding experimental uncertainties. Fig. \ref{SI Fig4} c-d show typical Raman images plotting the Si-Si peak shift for a virtual substrate following the CMP step. The most salient feature in the images is the familiar crosshatch pattern of strain fluctuations  running along (110) directions \cite{Sawano2003, Fitzgerald1997,Hsu1992}. Several prior works have shown that the Si-Si LO$_z$ peak, $\omega$, in the Si$_{1-x}$Ge$_{x}$ alloy shifts with composition, $x$, and strain, $\epsilon_{||}$, by comparable amounts \newline

$\omega = 520.7 - 66.9x - 730(70)\epsilon_{||}$,\cite{10.1063/1.2913052} \newline

$\omega = 520.5 - 62x - 815\epsilon_{||}$,\cite{Tsang1994} \newline

$\omega = 520 - 70.5x - 830\epsilon_{||}$,\cite{Perova2011}, or \newline

$\omega = 521 - 62x - 845\epsilon_{||}$.\cite{Rouchon2014} \newline

For the composition, $x=0.30\pm0.01$, from the SIMS data in Fig. \ref{SI Fig2} a, we calculate a expected peak position as 500.6 ± 0.7. This is slightly larger than the range (498.5-499.5 cm$^{-1}$) of measured mean Si-Si peak positions, Fig. \ref{SI Fig4} f, which is consistent with small average tensile strain ($\epsilon_{||} \sim$1.5-3 $\times10^{-3}$) in all substrates, Fig. \ref{SI Fig4} g. In addition, the standard deviations $\sim$0.2 cm$^{-1}$ of the peak distributions for all substrates indicates strain fluctuations on the order of $\delta \epsilon_{||}=3\times10^{-4}$, which is similar to other prior works on comparable materials processes.

\begin{figure*}[h]
\includegraphics[width=1.0\textwidth]{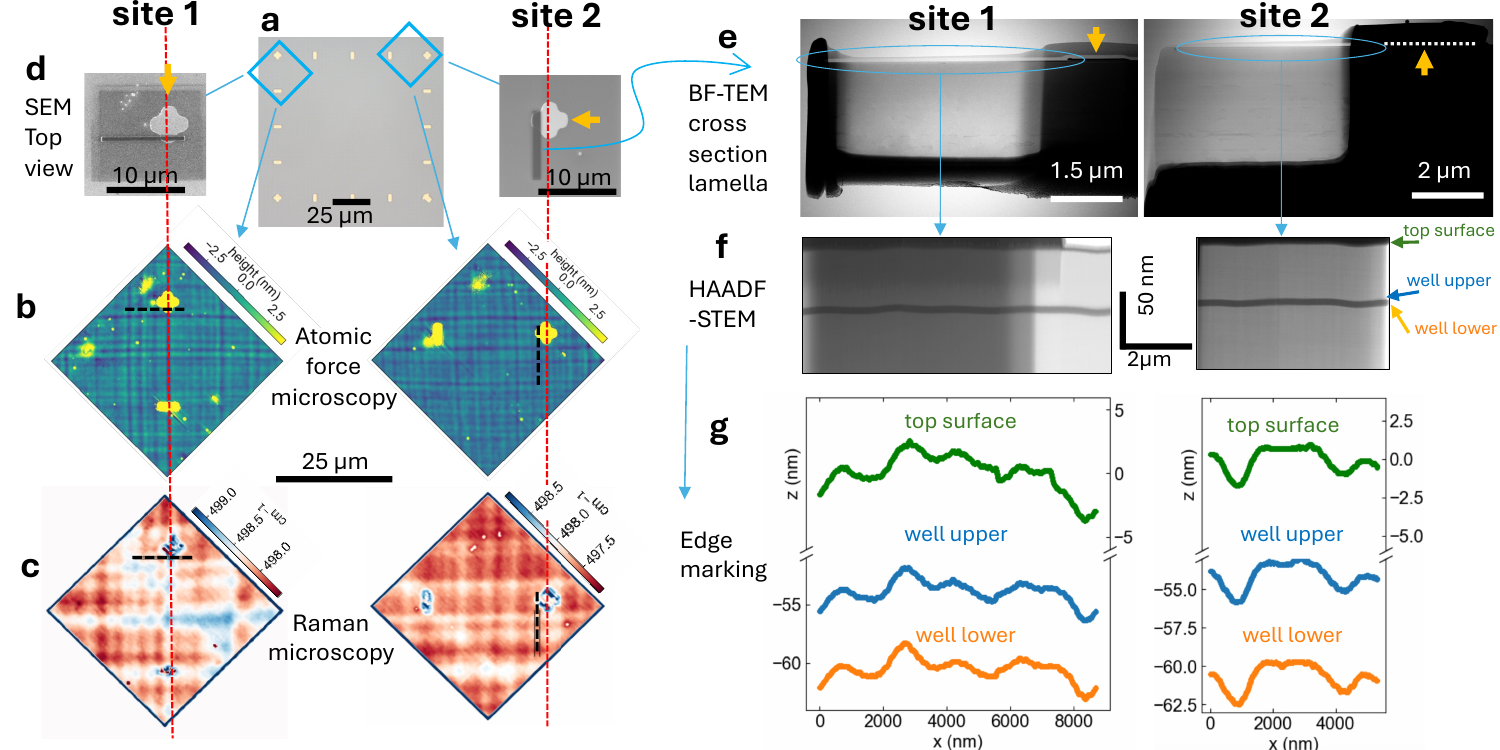}
\caption{\textbf{Process for spatially-aligned imaging to measure correlations between surface and interface structure, and strain fluctuation.} \textbf{a} Metal alignment marks (50 nm-thick Au/ 5 nm-thick Ti) are microfabricated using a liftoff process. Near sites 1 and 2 (blue squares), we perform multi-perspective imaging aligned to `+'-shaped marks.  \textbf{b} Surface topographies for sites 1 and 2 are measured using atomic force microscopy (AFM) imaging. \textbf{c} Next, the near-surface strain variation (Si-Si $LO_z$ phonon shift) is measured using Raman microscopy imaging. Red dashed lines running across the panels indicate the spatial alignment between features. \textbf{d} Then cross-sectional lamella are cut-out using standard focused ion beam (FIB) and lift-out techniques. Scanning electron microscopy (SEM) images show the top surface views of the FIB sites (darkest regions). The corresponding location is indicated by black dashed lines in the AFM and Raman images. \textbf{e} cross-sectional views of the lamella in bright-field (BF) TEM. Blue ovals indicate the strained-Si well region, orange arrows point out the cross-sectioned metal alignment features location which is verified in cross-sectional SEM imaging with sub-100-nm precision (not shown). \textbf{f} High-angle annular dark field scanning transmission electron microscopy images of the quantum well region. Note that vertical axis is exaggerated to 32:1 scale to emphasize the well structure, as indicated by the scale bars. \textbf{g} A Canny edge marking is used to mark and plot the top surface and the the s-Si quantum well upper and lower interfaces.}
\label{SI Fig5} 
\end{figure*}

\begin{figure*}
\centering
\includegraphics[width=6 in]{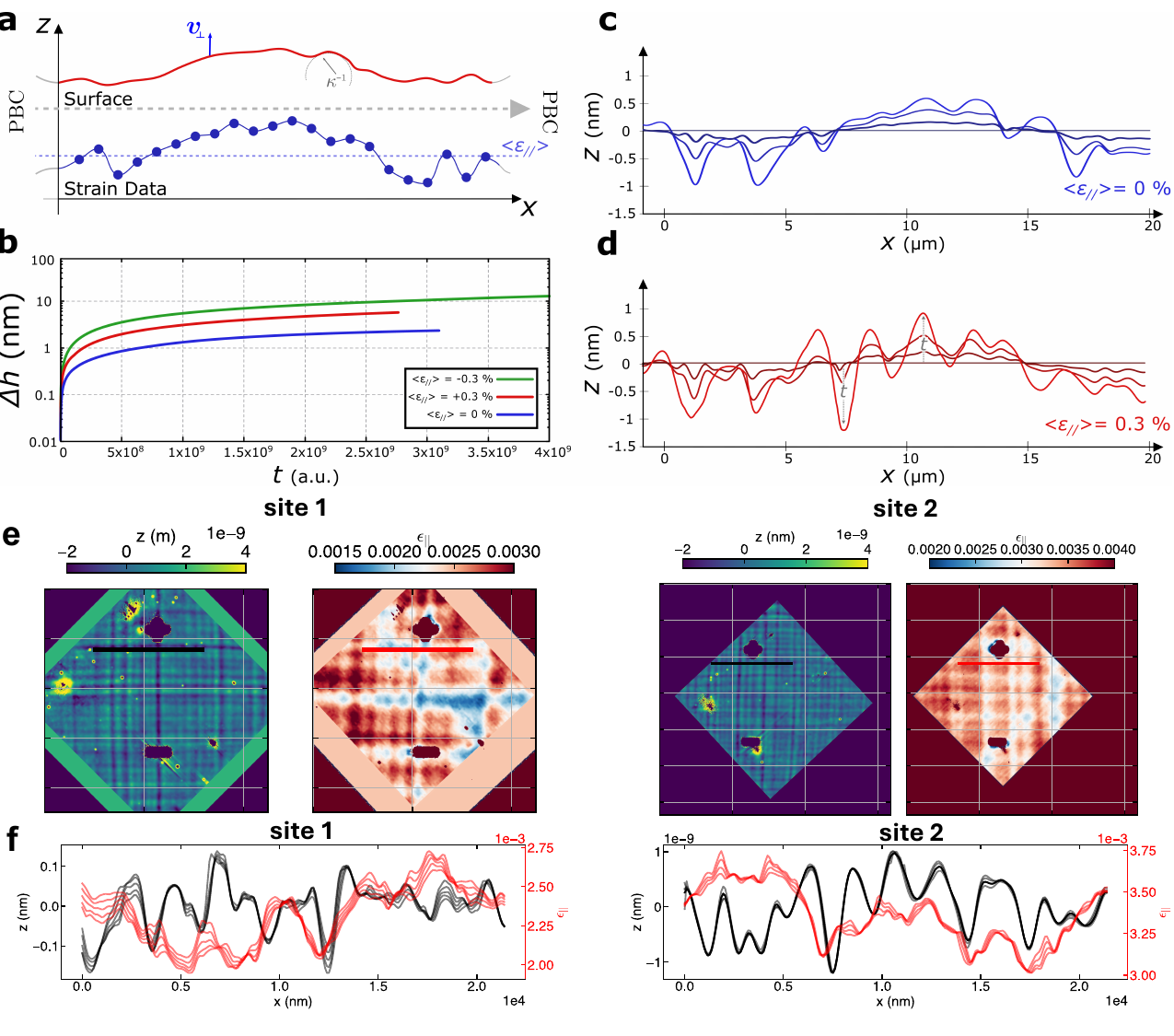}
\caption{\textbf{The simulation framework and its outcomes, including surface roughness evolution under varying in-plane strain and representative surface profiles at different time steps.} \textbf{a} Schematic representation of the model \textbf{b} Surface roughness (difference between maximum and minimum of the surface profile) over the simulation time for different values of the average in-plane strain. \textbf{c} Snapshots of the surface profiles obtained at different simulation times for the case of $\langle\epsilon_{\parallel}\rangle = 0\%$. The starting condition at $t=0$ corresponds to the flat profile. \textbf{d} Snapshots of the surface profiles obtained at different simulation times for the case of $\langle\epsilon_{\parallel}\rangle = 0.3\%$. The starting condition at $t=0$ corresponds to the flat profile. \textbf{e} Site 1 (left) and site 2 (right) AFM an Raman microscopy images showing the locations, indicated by black and red lines of length 21 µm, for \textbf{f} the strain, $\epsilon_{\parallel}$, and topography, z, used in the manuscript Fig. 3. These line are near the sites for cross-sectional HAADF-STEM imaging, just about 2 µm offset to avoid alignment marks.} 
\label{SI Fig6}
\end{figure*}

\begin{figure}
\includegraphics[width=1.0\textwidth]{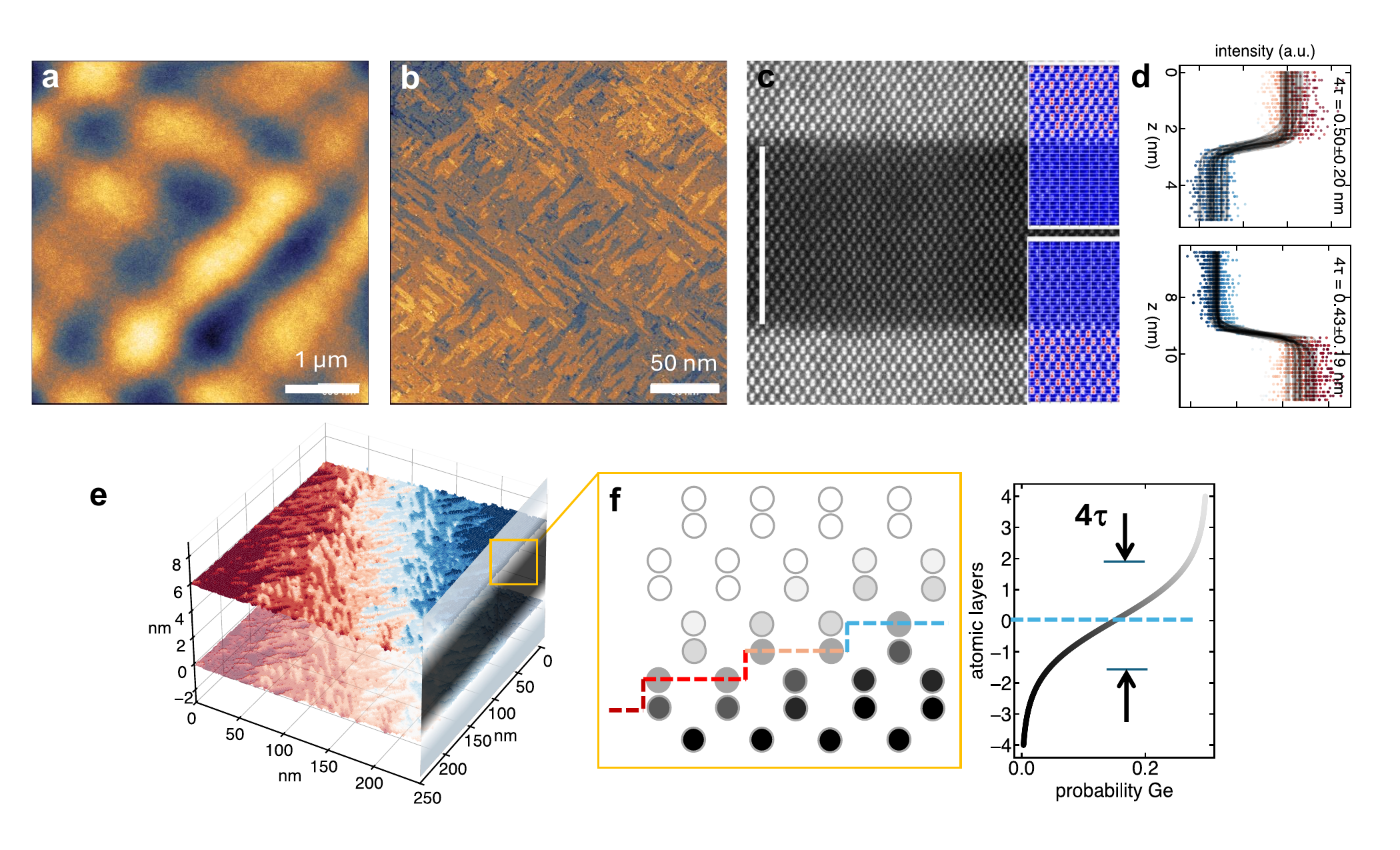}
\caption{\textbf{The basic ingredients of our atomic structure model used in our calculations of the influence of atomic-step and alloy disorder on valley splitting variability.} \textbf{a} Atomic resolution scanning tunneling microscopy (STM) data showing the cross-hatch surface of the virtual substrate after the pregrowth anneal (wafer \#7). \textbf{b} Higher-magnification STM showing the cascade of atomic steps on the surface. \textbf{c} Quantum well interfaces in a high-angle annular dark field scanning transmission electron microscopy image of a completed heterostructure (wafer \#15).
The colored insets show the marking of image atomic columns to detect the column intensity, a measure of the composition, in the interface region. \textbf{d} The estimation of the interface widths using a sigmoid fit to the column intensity along each atomic plane across the interface. The metric for the interface width is $4\tau$, where $\tau$ is the sigmoid width parameter. We find $4\tau=0.47\pm0.21$~nm. \textbf{e} The marked atomic planes in the STM data of a typical sloped region ($0.1^{\circ}$ local miscut) of the cross-hatch pattern. \textbf{f} To compile an atomic structure model of the well-region including both the atomic steps and alloy distribution across each interface, the effective local mean position of each interface is set by the atomic layers marked in the STM data, separated by 6.1 nm well width, then a specific alloy disorder realization (distribution of Ge around each mean local interface center) is calculated as a Bernoulli trial at each atomic site drawing from a sigmoid probability distribution with the width parameter, $4\tau=0.47\pm0.21$~nm, from the HAADF-STEM intensity and the center position at the local layer height from the STM.}
\label{SI Fig7} 
\end{figure}


\clearpage

\bibliography{bibliography.bib}  